\documentclass[a4paper,11pt]{article} 
\usepackage{jheppub} 
\usepackage{lineno} 
\usepackage{xcolor} 
\usepackage[T1]{fontenc} 
\usepackage{lmodern} 
\usepackage[english]{babel} 
\usepackage{csquotes} 
\usepackage{amsmath,amssymb,amsfonts} 
\usepackage{orcidlink} 
\usepackage{microtype}

\def\sj{\mathsf{j}}
\def\sa{\mathsf{a}}
\def\sT{\mathsf{T}}
\def\sL{\mathsf{L}}

\def\se{\mathrm{e}}
\def\so{\mathrm{o}}
\def\vt{\theta}

\def\mT{\mathcal{T}}
\def\mJ{\mathcal{J}}
\def\mDJ{\mathcal{DJ}}
\def\mDT{\mathcal{DT}}
\def\Ai{\scriptscriptstyle{\text{Airy}}}

\def\Tr{\mathrm{Tr}\hskip 1pt}

\renewcommand{\geq}{\geqslant}

\def\NS{\mathrm{\scriptscriptstyle N}}
\def\R{\mathrm{\scriptscriptstyle R}}

\def\bra#1{\left\langle\,#1\,\right|}
\def\ket#1{\left|\,#1\,\right\rangle}

\DeclareMathOperator*{\Res}{Res}

\title{Super Quantum Airy Structures and Matrix Models: \\ A Supercurrent Approach}

\author[a]{Leszek Hadasz\,\orcidlink{0000-0002-8142-8185}} 
\author[a,b]{Mykhailo Hontarenko\,\orcidlink{0009-0005-1354-7149}}

\affiliation[a]{Institute of Theoretical Physics, Jagiellonian University, ul. Łojasiewicza 11, Kraków, 30-348, Poland}
\affiliation[b]{Doctoral School of Exact and Natural Sciences, Jagiellonian University, ul. Łojasiewicza 11, 30-348 Kraków, Poland}

\emailAdd{leszek.hadasz@uj.edu.pl} \emailAdd{mykhailo.hontarenko@doctoral.uj.edu.pl}

\arxivnumber{xxxx.yyyyy \ \ ver. Aug.X, 2026}

\abstract{We develop a super-current formulation of $\mathcal{N}=1$ super-Virasoro constraints for external-source models. By assigning bosonic and fermionic monodromies independently, we unify the NS--NS, NS--R, R--NS, and R--R sectors. For each, a super-Miwa transformation represents the projected super energy-momentum tensor as a differential operator in spectral variables. Dilaton shifts yield Super Quantum Airy Structures in the NS--NS and R--NS sectors, whereas the R--R and NS--R sectors require an additional odd coordinate. Our principal result is an exact similarity transformation relating the NS--NS and R--R differential constraints. The intertwining prefactor combines a bosonic determinant, a cubic Airy weight, and a Grassmann exponential kernel defined by the difference of Neveu--Schwarz and Ramond fermionic propagators. These results establish a concrete candidate system of Ward identities for a 
supersymmetric extension of the Kontsevich model, delineating the precise 
algebraic and analytic properties of the measure and fermionic kernel required 
for its potential matrix-model realization.}

\begin{document}
\maketitle
\flushbottom

\section{Introduction}

The Kontsevich--Witten theorem provides one of the fundamental links between
two-dimensional quantum gravity, intersection theory, integrable systems and
matrix models \cite{Witten91,Witten90,Kontsevich92}.  It identifies the
generating function of intersection numbers on the moduli spaces of stable
curves with the partition function of the matrix Airy model.  The ribbon-graph
expansion of this integral is closely related to a cell decomposition of the
moduli space \cite{Zvonkine04}; see also \cite{Zvonkine12,GL24} for
introductions to the geometry of moduli spaces and to the Kontsevich
construction.  This correspondence has motivated a broad range of developments
in algebraic geometry, topological quantum field theory, matrix models and
topological recursion \cite{EO07,ABCO24}.

The external matrix of the Kontsevich model is not a formal device.  In the
topological string interpretation, the B-model on a local Calabi--Yau
threefold $zw-H(p,x)=0$ is governed by the spectral curve
$\Sigma=\{H(p,x)=0\}$ together with the differential $p(x)\,\mathrm{d}x$,
locally represented by a chiral boson through $p(x)=\partial_x\phi(x)$; the
modes of this field describe deformations of the complex structure at
infinity.  Two complementary brane descriptions of the same closed-string
amplitudes then produce two matrix models \cite{ADK03,DV02}.  Compact B-branes
give an ordinary Hermitian matrix model, whose back-reaction is absorbed into
the closed-string geometry in the geometric transition to the mirror manifold \cite{CIV01,DV02}.
Non-compact B-branes instead remain at finite distance, their positions being
encoded in the eigenvalues of an external $N\times N$ matrix; integrating out
the corresponding open strings produces a rank-$N$ Kontsevich-type integral
\cite{ADK03,GR03}, which in the Airy case is the original Kontsevich model.
The bridge between the closed-string couplings and this external-source
description is the Miwa transformation \cite{Miwa82}.  Denoting by $l_i$ the
eigenvalues of the external matrix $L$, it reads, in the normalisation used
below,
\begin{equation}
t_m\;\equiv\;g_m\;=\;\frac{1}{m}\sum_{i}l_i^{-m}\;=\;\frac{1}{m}\Tr L^{-m}.
\label{eq:intro-Miwa}
\end{equation}
For finite $N$, this transformation packages the infinitely many
hierarchy times $t_m$ into $N$ spectral variables $l_i$ and restricts
the theory to a finite-rank Miwa locus.  In the formal limit
$N\to\infty$ used below, the same relation becomes an
infinite-dimensional change of variables and recovers the full hierarchy
of times.  In both descriptions it turns the Virasoro or, more generally,
the $W$--constraints \cite{Mikhailov93,MMMR21} into differential
equations in the external spectral variables.  Thus, the Miwa variables
provide both the finite-$N$ external-source parametrisation and its
formal infinite-dimensional extension.

A foundational framework for the supergeometric extension of mirror symmetry was 
established by Aganagic and Vafa \cite{Aganagic:2004yh}, who developed techniques for obtaining 
the mirrors of Calabi-Yau supermanifolds as super Landau-Ginzburg theories. By 
implementing T-duality for fermionic coordinates, they demonstrated that super-Calabi-Yau 
targets, such as the twistorial $\mathbb{C}P^{3|4}$, possess mirror duals whose 
complex structure depends explicitly on the K\"ahler parameter $t$. These findings are 
highly relevant to the current construction, as they provide the string-theoretic 
underpinnings for the B-model on supermanifolds where the modes of the super-current 
$\mJ(x, \theta)$ describe deformations of the complex structure at infinity. Vafa's 
construction of the mirror superpotential, which balances bosonic and fermionic 
contributions to preserve the central charge, serves as a conceptual precursor to our 
derivation of the super-Kontsevich prefactor and the resulting intertwining relations 
between the Neveu--Schwarz and Ramond sectors.

It is therefore natural to ask whether one can extend relations outlined
in the first paragraph to supersymmetric
theories and  invariants of supergeometric moduli spaces
\cite{AM91,Takama92,BB93,AIMZ92,Yost92,Plefka95,Eberhardt26}.  The geometric setting
suggested by supersymmetry is the theory of super-Riemann surfaces
\cite{Witten15,Witten19}, whose moduli spaces are substantially more
complicated than the ordinary $\overline{\mathcal M}_{g,n}$.  In particular,
the non-projectedness theorem of Donagi and Witten shows that for sufficiently
large genus the supermoduli space cannot be holomorphically projected onto its
reduced bosonic space \cite{DW15}, so that it cannot in general be represented
globally as the total space of a purely odd vector bundle over the moduli
space of ordinary curves.  The inclusion of Neveu--Schwarz and Ramond
punctures introduces further geometric and analytic complications; see
\cite{Witten19} for a detailed discussion.

The reduced geometric data of a split super-Riemann surface include a spin
structure, that is, a choice of a square root $K_\Sigma^{1/2}$ of the
canonical bundle, which leads naturally to moduli spaces of spin curves.  On
these spaces Norbury constructed cohomological classes whose pushforwards
define the classes $\Theta_{g,n}$ on $\overline{\mathcal M}_{g,n}$
\cite{Norbury21,Norbury25}.  Their generating function is described by the
Br\'ezin--Gross--Witten partition function and satisfies the corresponding
Virasoro constraints \cite{Norbury25}.  This provides an important example in
which structures motivated by supergeometry give rise to a concrete system of
differential constraints on an ordinary moduli space.

The algebraic framework for organising such systems of constraints is provided
by Quantum Airy Structures \cite{KS17,ABCO24,BCJ22}.  An Airy structure
consists of differential operators closing under commutation and admitting a
recursively determined partition function, whose free energy generates the
correlators or geometric invariants of the corresponding theory.  This
formalism contains the Witten--Kontsevich model as a special case and is
closely related to topological recursion; see \cite{Bouchard24} for a
comprehensive introduction.  The construction was extended to Super Quantum
Airy Structures in \cite{BCHORS20}, where bosonic and fermionic differential
operators are combined into a representation of a Lie superalgebra; suitable
Super Quantum Airy Structures reproduce the invariants associated with the
Br\'ezin--Gross--Witten model and the Norbury classes.

A complementary perspective comes from conformal field theory.  In the bosonic
setting, matrix-model loop equations and topological recursion can be
formulated in terms of a free bosonic current and an appropriate state at
infinity \cite{Kostov10,KO10}.  The choice of this state encodes the potential
and the polarisation, including the Bergman kernel, while its local behaviour
is tied to the branch points of the underlying spectral curve.  From this
viewpoint the Virasoro constraints arise from the singular part of the
energy-momentum tensor, and the Airy structure is obtained by conjugating the
free-field representation by the operators implementing the dilaton shift and
the change of polarisation.

Several supersymmetric extensions of topological recursion have been proposed.
The $\mathcal N=1$ super-topological recursion developed in
\cite{BO18,BO21,Osuga19,Osuga22} is motivated by super-eigenvalue models
\cite{McArthur93,ABBEM93,AIMZ92,BB93}.  Its initial data consist of polarised
bosonic and fermionic differentials on a local super-spectral curve, and the
resulting recursion is algebraically equivalent to a class of Super Quantum
Airy Structures \cite{BCHORS20}.  More recently, superconformal topological
recursion was introduced from a different, explicitly geometric perspective
\cite{AKOO25}.  It is formulated directly in terms of superconformal geometry
and of the fundamental super-bidifferential on a super-Riemann surface; for
split super-Riemann surfaces this bidifferential combines the roles of the
Bergman and Szeg\H{o} kernels \cite{DP15,Witten15,Witten19}.  Its relation to
Super Quantum Airy Structures is described through partial super-Airy
structures.

Against this background, the relation between abstract super-Virasoro modes, bosonic and fermionic spectral variables, and matrix-model
differential equations remains less explicit than in the ordinary Kontsevich
theory.  The obstruction is concentrated in a single object.  A supersymmetric
extension of the Miwa transformation \eqref{eq:intro-Miwa} must distinguish
the Neveu--Schwarz and Ramond sectors, incorporate the fermionic zero mode in
the Ramond representation, and determine the Grassmann-valued deformation of
the Kontsevich prefactor.  Supplying these three ingredients is the purpose of
the present work.

We address them using a unified super-current formalism.  We combine the free
bosonic and fermionic currents into a single super-current $\mJ(x,\vartheta)$
and construct the corresponding super energy-momentum tensor.  Assigning
monodromies to its bosonic and fermionic components independently leads to
four sectors, which we denote by NS--NS, NS--R, R--NS and R--R; throughout,
the first label refers to the monodromy of the bosonic current and the second
to that of the fermionic one, so that, for instance, R--NS stands for
$\mJ=\vt\jmath_\R+\psi_\NS$.  This formulation allows the corresponding
super-Virasoro constraints to be derived from the same residue construction in
all four sectors.

Our main results can be summarised as follows.

\medskip
\noindent First, for each of the four sectors we introduce a
sector-dependent super-Miwa transformation expressing the bosonic and
fermionic times in terms of commuting spectral parameters $l_i$ and their
Grassmann partners $\theta_i$.  These transformations are fixed by the
requirement that the annihilation part of the super-current be represented by
the covariant superderivative
$D_i=\partial_{\theta_i}+\theta_i\partial_{l_i}$, and we use them to obtain
the projected super energy-momentum tensor as an explicit 
differential operator in $(l_i,\theta_i)$, computing all contributions
sector by sector.  In the R--NS and NS--R sectors the relevant constraint is
the one associated with the maximal non-anomalous subalgebra of the Ramond
algebra containing the positive modes, spanned by $\{G_{\geqslant0},
L_{\geqslant1}\}\cup\{L_0-\tfrac1{16}\}$.

\medskip
\noindent Second, we construct the corresponding Airy operators by
applying sector-dependent dilaton shifts to the projected tensors.  A
single shift of the bosonic current produces simultaneously the linear
terms in both the Virasoro and  super-Virasoro generators; this is a
direct consequence of working with one super-current rather than with
$T(x)$ and $G(x)$ separately.  In the NS--NS and R--NS sectors the shift
of the lowest positive bosonic mode supplies leading derivatives with
respect to all bosonic and fermionic variables, and the constraints form
ordinary Super Quantum Airy Structures.  When the fermionic current is in
the Ramond sector, its zero mode $\psi_0 = \frac{q_0}{2} + \frac{\partial}{\partial q_0}$ contains both multiplication and differentiation and therefore requires separate treatment.  In the R--R sector the odd generators are half-integer, so that no $G_0$ occurs: after the cubic dilaton shift the fermionic leading derivatives start with $\partial_{q_1}$, while $q_0$ remains one additional odd coordinate in the sense of \cite{BCHORS20}.
The NS--R sector is the only one in which the Ramond zero mode and the
lowest odd generator $G_0$ occur simultaneously.  Shifting $\sa_0$
produces the full zero-mode combination
$\psi_0=\partial_{q_0}+q_0/2$ rather than a pure leading derivative, and
we therefore discuss in Section~\ref{sec:airy} two possible Airy
realisations.  The first excludes $G_0$ after shifting $\sa_0$.  The
second retains $G_0$ by shifting $\sa_1$ and supplementing the resulting
family with the auxiliary central constraint
$\sa_0=\partial_{g_0}$.  The two choices are therefore correlated through
the choice of the dilaton shift and the treatment of the lowest
generators.  In both cases $q_0$ remains the single additional odd
coordinate.
 
We also formulate the more general deformation obtained by combining the
dilaton shift with a change of bosonic and fermionic polarisation,
governed by coefficients $\varphi_{mn}$ and $\chi_{rs}$; the explicit
operators studied here correspond to trivial additional polarisation,
$\varphi_{mn}=\chi_{rs}=0$, and to a single non-vanishing dilaton moment
$\mu_k$ at a time.  At the level of the local quadratic density we compare
the resulting super-current construction with the $\mathcal N=1$
super-topological recursion of \cite{BO21} and with the super quadratic
Casimir underlying the superconformal topological recursion of
\cite{AKOO25}. 

\medskip
\noindent Third, on the Miwa locus we establish an exact
similarity transformation between the differential realisations of a
modified NS--NS constraint and, first, the unshifted R--R constraint and,
after adding the cubic weight, the R--R Airy constraint with one additional
odd variable.
Writing $l_i=\lambda_i^2$, we first introduce the sector-changing prefactor
\begin{equation}
F_0(\lambda,\theta)
=
\prod_{i=1}^{N}\lambda_i^{-1}
\prod_{i<j}(\lambda_i+\lambda_j)^{-2}
\exp\left[
-\frac{1}{2}
\sum_{i<j}
\theta_i\theta_j\,
\frac{\lambda_i-\lambda_j}
{\lambda_i\lambda_j(\lambda_i+\lambda_j)}
\right],
\label{eq:intro-sector-prefactor}
\end{equation}
whose Grassmann kernel is precisely the difference between the
Neveu--Schwarz and Ramond fermionic two-point functions evaluated at the Miwa
points.  It intertwines the unshifted NS--NS and R--R operators,
\begin{equation}
\mT_{\R\R}^{\geqslant-\frac12}(l_i,\theta_i)
=
F_0^{-1}\,
\mT_{\NS\NS}^{\geqslant-\frac12}(l_i,\theta_i)\,
F_0 .
\label{eq:intro-sector-intertwiner}
\end{equation}
The full super-Kontsevich prefactor is obtained by adjoining the cubic Airy
weight,
\begin{equation}
F(\lambda,\theta)
=
F_0(\lambda,\theta)\,
\exp\left[-\frac{\mu}{3}\sum_i\lambda_i^{3}\right],
\label{eq:super-Kontsevich-prefactor}
\end{equation}
in terms of which the exact operator identity reads
\begin{equation}
F^{-1}
\left[
4\mT_{\NS\NS}^{\geqslant-\frac12}(\lambda_i^2,\theta_i)
-\mu^2\theta_i\lambda_i^2
+\mu q_0
-2\mu\nabla_{q_0}
\right]
F
=
{4\mT}^{\geqslant - \frac12}_{\R\R,\Ai}(\lambda_i^2,\theta_i),
\label{eq:intro-conjugation}
\end{equation}
where the Ramond Airy operator is
\begin{equation}
{4\mT}^{\geqslant - \frac12}_{\R\R,\Ai}(\lambda_i^2,\theta_i)
=
4\mT_{\R\R}^{\geqslant-\frac12}(\lambda_i^2,\theta_i)
-
2\mu
\left(
\theta_i\frac{\partial}{\partial\lambda_i}
+
\lambda_i\frac{\partial}{\partial\theta_i}
+
\frac{\partial}{\partial q_0}
\right).
\label{eq:intro-RR-Airy}
\end{equation}
Here $\nabla_{q_0}=\psi_0+S(\lambda,\theta)$ is the covariant Ramond
zero-mode operator; the odd multiplication term $S(\lambda,\theta)$ is
determined by the Grassmann part of $F$ and is required because $F$ does not
commute with $\partial_{q_0}$. These identities act on
functions of the spectral variables $(\lambda_i,\theta_i)$.  They do not
by themselves define an isomorphism of the abstract NS--NS and R--R
constraint algebras: $q_0$, $\psi_0$ and $\nabla_{q_0}$ are Ramond-sector
objects, while the factors entering $F_0$ and $F$ are not manifestly formal
power series in the abstract Miwa times.  Consequently, no relation between
the abstract partition functions is asserted away from the Miwa locus.
Thus the four monodromy assignments remain distinct at the current-algebra
level, although the NS--NS and R--R differential realisations become
equivalent after restriction to the Miwa locus.      Thus $F_0$ performs the change of sector,
while the cubic factor generates the compatible bosonic and fermionic Airy
shifts.  We compute the transformed operator in closed form, including its
first-order differential coefficients and all linear and cubic Grassmann
contributions to its zeroth-order part. 

Finally, inspired by the matrix-derivative approach to the Kontsevich and
supersymmetric eigenvalue models \cite{Mikhailov93,CM92,Plefka95}, we use
the standard radial matrix-derivative identity to identify the bosonic
part of our operators; in the purely bosonic limit it reproduces the
Gross--Newman-type matrix differential equation associated with the
Kontsevich integral.  We do not, however, interpret the super-Miwa
operators as radial operators of an underlying supermatrix integral,
since no established external-source construction reproduces the four
sector-dependent operators derived here together with their Ramond
zero-mode realisation.  What the present construction provides instead is
a concrete system of Ward identities that any future
super-Kontsevich-like model would have to reproduce, together with
explicit constraints on its measure and on its fermionic kernel; we
return to this in Section~\ref{sec:outlook}.

The paper is organised as follows.  In Section~\ref{sec:section2} we
review free bosonic and fermionic fields on $\mathbb{CP}^{1}$ and
construct the super-current in the Neveu--Schwarz and Ramond sectors.  We
then combine the independent monodromy assignments of its bosonic and
fermionic components into the NS--NS, NS--R, R--NS and R--R sectors.  In
Section~\ref{sec: Section 3} we introduce the corresponding
sector-dependent super-Miwa transformations and derive the differential representations of the projected super energy-momentum tensor.  In Section~\ref{sec:airy} we construct the
simplest sector-dependent Airy operators by means of dilaton shifts and
discuss the two possible realisations in the NS--R sector; we also
consider a general change of polarisation and compare the local
super-current construction with $\mathcal N=1$ super-topological
recursion and with superconformal topological recursion.  In
Section~\ref{sec:matrix} we recall the radial matrix-derivative identity
and its relation to the external-source Kontsevich integral and to ribbon
graphs, and then construct the prefactor intertwining the NS--NS and R--R
sectors and establish the corresponding similarity transformation for the
R--R Airy constraint.  Section~\ref{sec:outlook} discusses the remaining
steps towards a super-Kontsevich-like matrix-model realisation and the
related open problems.

\section{Free fields on the complex plane}
\label{sec:section2}

\subsection{Boson field}
A free, massless, bosonic field on a Riemann sphere, with local coordinates centered at a point $p$ denoted by $z,$ is a generating function,
\begin{equation}
j(z) = \sum\limits_{m\in\mathbb Z} \frac{a_m}{z^{m+1}},
\end{equation}
for a set of elements which, together with an identity, form a complex Heisenberg algebra  with the bracket (commutator)
\begin{equation}
\label{eq:bosonic_Heisenberg_algebra}
[a_m,a_n] = m\delta_{m+n,0}\hbox{\boldmath$1$}.
\end{equation}
Note for the future reference that this algebra has an obvious, infinite group of automorphisms, 
\begin{equation}
\label{eq:Heisenbarg_algebra_authomorphism}
    a_m \to a_m + \mu_m, \qquad \mu_m \in \mathbb{C}.
\end{equation}
Here and in what follows we no longer explicitly write the unit operator $\hbox{\boldmath$1$}$ since its presence is always clear from the context.
The simplest realization of the Virasoro algebra 
\begin{equation}
\label{eq:Virasoro_section_2}
[L_m,L_n] = (m-n)L_{m+n} + \frac{c}{12}m\left(m^2-1\right)\delta_{m+n,0}
\end{equation}
can be constructed in the universal enveloping algebra of the Heisenberg algebra as  modes 
\begin{equation}
\label{eq:Virasoro_algebra_generators_1}
L_m = \Res\limits_{z=0} z^{m+1}T(z)
\end{equation}
of the local field 
\begin{equation}
T(z) = \frac12\lim\limits_{w \to z}\left(j(w)j(z) - \frac{1}{(w-z)^2}\right) \equiv \frac12 :\!j(z)j(z)\!:.
\label{eq:T(z)}
\end{equation}
Here and in what follows double dots denote  the ``creation-annihilation'' normal ordering operator,
\begin{equation}
\label{eq:normal_ordering_bosonic_modes}
:\!a_ma_n\!:\; = \left\{\begin{array}{rcl}
     a_na_m &&  m \geqslant 0,\,  n < 0
     \\
     a_m a_n && \mathrm{otherwise}.
\end{array}
\right.
\end{equation}
Since
\begin{equation}
\label{eq:TT_product}
T(z)T(w) 
=
\frac12\frac{1}{(z-w)^4} + \frac{1}{(z-w)^2} :\!j(z)j(w)\!: + :\!T(z)T(w)\!:
\end{equation}
the modes of $T(z)$ above satisfy (\ref{eq:Virasoro_section_2}) with $c=1.$ 

It is often sufficient -- in particular in computing various commutation relations -- to retain only the information about singular (in the limit $z\to w$) terms in products of local operators. One thus typically presents formulas like (\ref{eq:TT_product}) in the form of the Operator Product Expansion (OPE)
\begin{equation}
\label{e:TT_OPE}
T(z)T(w)  \sim \frac{c}{2}\frac{1}{(z-w)^4} + 
\frac{2 T(w)}{(z-w)^2} + \frac{\partial T(w)}{z-w}, \qquad c = 1,
\end{equation}
where the $\sim$ symbol means ``up to terms finite in the limit of coinciding arguments''. In particular
\begin{equation}
\label{eq:Tj_OPE}
T(z) j(w) \sim \frac{j(w)}{(z-w)^2} + \frac{\partial j(w)}{z-w}.
\end{equation}
In the CFT language the content of this equation is rephrased as a statement that $j(z)$ is primary fields of scaling dimension 1. Equivalently, for a holomorphic change of coordinates $x = x(z)$ the field $j(z)$ transform as
\begin{equation}
\label{eq:j_transformation}
j'(x)dx = j\big(x(z)\big)dx(z).
\end{equation}

Using solely the free field $j(z)$ we can in fact generate a whole family of ``energy-momentum tensors''
\begin{equation}
T^Q(z) = T(z) + \frac{Q}{2}\partial j(z),
\end{equation}
whose modes satisfy the Virasoro algebra with $c=1-3Q^2.$ Notice however that the $T^Q\,j$ OPE possess a third order pole with a residue proportional to $Q:$
\begin{equation}
\label{eq:TQj_OPE}
T^Q(z) j(w) \sim -\frac{Q}{(z-w)^3} + \frac{j(w)}{(z-w)^2} + \frac{\partial j(w)}{z-w}
\end{equation}
and, unless $Q=0,$ $j(z)$ does not behave covariantly under holomorphic transformations generated by $T^Q.$

A distinguished class of Virasoro algebra representations is formed by the highest weight reps, characterized by the existence of the (cyclic) highest weight vector $\nu_\Delta,\ \Delta \in \mathbb{C},$ satisfying
\begin{equation}
L_0\nu_\Delta = \Delta\,\nu_\Delta, \qquad L_n\nu_\Delta = 0, \; n > 0.
\end{equation}
The Fock vacuum $\ket{0}$ with $a_n\ket{0} = 0,\; n \geqslant 0$ is a highest weight vector of (\ref{eq:Virasoro_algebra_generators_1}), satisfying
\begin{equation}
L_n\ket{0} = 0, \qquad n \geqslant -1.
\end{equation}
Let $\Sigma$ be a Riemann surface \cite{Miranda95} and  consider a holomorphic map $\mathbb{CP}^1 \to \Sigma.$ Denote by $x$  a local coordinate on $\Sigma$ centered at the image of $p.$ Suppose now that this point is a (second order) branch point, so that locally $x = z^2.$ A closed path $x \to \mathrm{e}^{2\pi i}$ around $x =0$ on $\Sigma$ corresponds to $z \to -z$ on $\mathbb{CP}^1$ and we split $j(z)$ into components which behave uniformly under this map,
\begin{equation}
j(z) = j_\so(z) + j_\se(z), \qquad
j_\so(z) = \sum\limits_{m\in\mathbb Z}\frac{a_{2m}}{z^{2m+1}},\qquad j_\se(z) = \sum\limits_{m\in\mathbb Z}\frac{a_{2m+1}}{z^{2m+2}}.
\end{equation}
We have
\begin{equation}
j_\so(w)j_\so(z) =\; :\!j_{\so}(w)j_\so(z)\!: + \frac{2w z}{(w^2-z^2)^2}\;= \; \frac{1}{2}\frac{1}{(w-z)^2} + :\!j_{\so}(z)j_\so(z)\!: - \frac{1}{8z^2} + \mathcal{O}\big(w-z\big)
\end{equation}
and, consequently, modes of 
\begin{equation}
\label{eq:To_tensor}
T_{\so}(z) = \lim\limits_{w\to z}\left(j_\so(w)j_\so(z) - \frac{1}{2}\frac{1}{(w-z)^2}\right) =\; :\!j_{\so}(z)j_\so(z)\!: - \frac{1}{8z^2}.
\end{equation}
satisfy the Virasoro algebra with central charge $c=1.$ Since
\begin{equation}
T_\so(w)j_\so(z) = \frac{4wz}{(w^2-z^2)^2}j(w) + :\!T_\so(w)j_\so(z)\!:\; \sim \; \frac{j_\so(z)}{(w-z)^2} + \frac{\partial j_\so(z)}{w-z},
\end{equation}
the field $j_\so$ transforms, in a theory with conformal transformations generated by $T_\so,$ as a primary field with scaling dimension 1 and thus
\begin{equation}
\jmath_{\NS}(x) = \frac{d\sqrt{x}}{dx}j_{\so}(z)\Big|_{z=\sqrt{x}} =\frac{1}{2\sqrt{x}}j_\so\left(\sqrt{x}\right) =  \frac12\sum\limits_{m\in \mathbb{Z}} \frac{a_{2m}}{x^{m+1}}.
\end{equation}
It is now useful to write in N (used here as an abbreviation for Neveu-Schwarz) sector
\begin{equation}
\label{eq:bosonc_current_on_x_plane}
\jmath_\NS(x) = \sum\limits_{m\in\mathbb Z}\frac{\sa_m}{x^{m+1}},\qquad \sa_m = \frac12 a_{2m}, \hskip 1cm [\sa_m,\sa_n] = \frac{m}{2}\delta_{m+n,0}\hbox{\boldmath$1$}.
\end{equation}
Since
\begin{equation}
\jmath_\NS(y)\jmath_\NS(x) =\; :\!\jmath_\NS(y)\jmath_\NS(x)\!: + \frac12\frac{1}{(x-y)^2}
\end{equation}
we have
\begin{equation}
\sT_\NS(x) = \lim\limits_{y\to x} \left(\jmath_\NS(y)\jmath_\NS(x)-\frac12\frac{1}{(y-x)^2}\right) =\; :\!\jmath_\NS(x)\jmath_\NS(x)\!:.
\end{equation}
The same result is also obtained if we directly transform $T_\so(z)$ using the formula
\begin{eqnarray}
\sT_\NS(x) = \left(\frac{dz}{dx}\right)^{2}\left(T_\so(z) -\frac{1}{12}\{x,z\}\right)_{z=\sqrt{x}}
=
\frac{1}{4x}\left(T_\so(\sqrt{x}) + \frac{1}{8x}\right),
\end{eqnarray}
Where $\{x,z\}$ denotes the Schwarzian derivative. Explicitly, splitting
\begin{equation}
\label{eq:jNS_splitting}
\jmath_\NS(x) = \jmath_{\NS,<}(x) + \jmath_{\NS,\geqslant}(x), \qquad \jmath_{\NS,<}(x) = \sum\limits_{m> 0}\sa_{-m} x^{m-1}, \qquad \jmath_{\NS,\geqslant}(x) = \sum\limits_{m\geqslant 0}\sa_m x^{-m-1},
\end{equation}
we have
\begin{equation}
\label{eq:TNS_splitting}
\sT_\NS(x) = \left( \jmath_{\NS,<}(x)\right)^2 + 2 \jmath_{\NS,<}(x)\jmath_{\NS,\geqslant}(x) + \left(\jmath_{\NS,\geqslant}(x)\right)^2.
\end{equation}
In what follows we will be interested in singular (for $x\to 0$) terms in $\sT_\NS(x).$ Using (\ref{eq:TNS_splitting}) and (\ref{eq:jNS_splitting}) we immediately get
\begin{equation}
\nonumber
\sT_{\NS}^{\geqslant -1}(x) 
\equiv  
\sum\limits_{m \geqslant -1}\frac{\sL_{\NS,m}}{x^{m+2}} 
=\left(\jmath_{\NS,\geqslant}(x)\right)^2 + 2 \Res_{y=0}\frac{1}{x-y} \jmath_{\NS,<}(y)\jmath_{\NS,\geqslant}(y),
\end{equation}
where we used the fact that, for $f(x)$ being a (formal) power series in $x,$ its sub-series containing negative powers of $x$ can be presented as:
\begin{equation}
\sum\limits_{m\geqslant 0}\frac{f_m}{x^{m+1}}
=
\Res\limits_{y=0} \sum\limits_{m\geqslant 0} \frac{y^m}{x^{m+1}} \frac{f_m}{y^{m+1}}
=
\Res\limits_{y=0} \sum\limits_{m\geqslant 0} \frac{y^m}{x^{m+1}} \sum\limits_n \frac{f_n}{y^{n+1}}
=
\Res\limits_{y=0} \frac{f(y)}{x-y}.
\end{equation}
Since $\jmath_{\NS,\geqslant}(x)\ket{0} = 0$ for all $x,$ we obviously have
\begin{equation}
\sT_{\NS}^{\geqslant -1}(x) \ket{0} = 0.
\end{equation}
The situation for the anti-periodic with respect to $x\to {\mathrm e}^{2\pi i}$ field $\jmath_{\se},$ whose modes generate the Ramond (or R) sector of the theory, is different. We have
\begin{equation}
j_\se(w)j_\se(z) =\; :\!j_\se(w)j_\se(z)\!:\; + \frac{w^2+z^2}{(w^2-z^2)^2} = \frac12\frac{1}{(w-z)^2}+ \; :\!j_\se(w)j_\se(z)\!:\; +  \frac{1}{8z^2} + \mathcal{O}\big(w-z)
\end{equation}
so that for the energy-momentum tensor
\begin{equation}
T_\se(z) = \lim\limits_{w\to z}\left(j_\se(w)j_\se(z) - \frac{1}{2}\frac{1}{(w-z)^2}\right) =\; :\!j_\se(w)j_\se(z)\!:\; +  \frac{1}{8z^2}
\end{equation}
we get
\begin{equation}
\sT_\R(x) = \left(\frac{dz}{dx}\right)^{2}\left(T_\se(z) -\frac{1}{12}\{x,z\}\right)_{z=\sqrt{x}}
=
\frac{1}{4x}\left(T_\se(\sqrt{x}) + \frac{1}{8x}\right)
=\; 
:\!\jmath_\R(x)\jmath_\R(x)\!:\; + \frac{1}{16x^2},
\end{equation}
with
\begin{equation}
\jmath_\R(x) = \frac{1}{2\sqrt{x}}j_\se\left(\sqrt{x}\right) = \frac12\sum\limits_{m\in\mathbb Z}\frac{a_{2m+1}}{x^{m+\frac32}}.
\end{equation}
Writing, similarly as for the $\jmath_{\NS}$ field,
\begin{equation}
\jmath_{\R}(x) = \sum\limits_{k\in\mathbb Z+\frac12}\frac{\sa_k}{x^{k+1}}, \qquad \sa_{k} = \frac12 a_{2k}, \qquad [\sa_k,\sa_l] = \frac{k}{2}\delta_{k+l,0}\,\hbox{\boldmath$1$},\quad k,l \in \mathbb{Z}+\frac12,
\end{equation}
and splitting the operator $\jmath_{\R}$ into modes with positive and negative indices,
\begin{equation}
\label{eq:je_splitting}
\jmath_\R(x) = \jmath_{\R,<}(x) + \jmath_{\R,>}(x), \qquad \jmath_{\R,<}(x) = \sum\limits_{k\geqslant\frac12}\sa_{-k} x^{k-1}, \qquad \jmath_{\R,>}(x) = \sum\limits_{k\geqslant \frac12}\sa_k x^{-k-1},
\end{equation}
we get
\begin{equation}
\label{eq:TR_splitting}
\sT_\R(x) = \left( \jmath_{\R,<}(x)\right)^2 + 2 \jmath_{\R,<}(x)\jmath_{\R,>}(x) + \left(\jmath_{\R,>}(x)\right)^2 + \frac{1}{16} \frac{1}{x^2},
\end{equation}
and in effect
\begin{equation}
\sT_\R^{\geqslant -1}(x) = \left(\jmath_{\R,>}(x)\right)^2 + 2\Res_{y=0}\frac{1}{x-y}\jmath_{\R,<}(y)\jmath_{\R,>}(y) + \frac{1}{16} \frac{1}{x^2} + \frac{1}{x}\sa_{-\frac12}^2,
\end{equation}
where the last term is a contribution from $ \left( \jmath_{\R,<}(x)\right)^2.$ Notice that $\sT_\R^{\geqslant -1}(x)$ does not annihilate the Fock vacuum,
\begin{eqnarray}
\sT_\R^{\geqslant -1}(x)\ket{0} = \Big(\frac{1}{16} \frac{1}{x^2} + \frac{1}{x}\sa_{-\frac12}^2\Big)\ket{0}.
\end{eqnarray}
\subsection{N=1 super-fields, super-current approach}
We now extend our studies to the case of $\mathcal{N} = 1$ super-conformal symmetry\footnote{Throughout the supersymmetric calculations, derivatives with respect
to Grassmann variables are understood as left derivatives.}. To this end one supplements the local, bosonic coordinate $z$ on $\mathbb{CP}^1$ with the Grassmann coordinate $\theta$ and requires
covariance of the constructed model with respect to the super-conformal transformations \cite{Witten19},
\begin{eqnarray}
\label{eq:SUSY_transformation}
\nonumber
z & \to & z' \equiv x  = f(z) + \theta\, \eta(z)\sqrt{f'(z)},
\\[-6pt]
\\[-6pt]
\nonumber
\theta & \to & \theta' \equiv \vt = \eta(z) + \theta\sqrt{f'(z)-\eta(z)\eta'(z)},
\end{eqnarray}
with an even function $f$ and an odd function $\eta.$ Any representation of (\ref{eq:SUSY_transformation}) consists of an even and an odd component; in particular, a bosonic, weight 1 representation of the conformal symmetry, i.e.\ the free scalar $j(z),$ can be combined with the ``free fermion'' $\chi(z)$ to form the $h=\frac12$ representation of (\ref{eq:SUSY_transformation}), the so called ``free super-current''
\begin{equation}
J(z,\theta) = \theta j(z) + \chi(z),
\end{equation}
where the (integer or half-integer) modes of $\chi(z) = \sum\limits_p \chi_p z^{-p-\frac12}$ satisfy the anti-commutation relation
\begin{equation}
\{\chi_p,\chi_q\} \equiv \chi_p\chi_q + \chi_q\chi_p = \delta_{p+q,0}, \qquad p,q \in \mathbb{Z} \quad \mathrm{or}\quad p, q \in \mathbb{Z}+\tfrac12.
\end{equation}
Under the super-conformal map (\ref{eq:SUSY_transformation}), the super-current $J$ transforms as
\begin{equation}
J'(x,\vt) = \big(D\vt\big)^{-2h} J(z,\theta), \quad h = \frac12, \qquad D = \frac{\partial}{\partial\theta} + \theta \frac{\partial}{\partial z}.
\end{equation}
For the special case of $ f(z) = z^2,\ \eta(z) = 0,$ we get
\begin{equation}
z = \sqrt{x}, \qquad \theta = \frac{1}{\sqrt{2}}x^{-\frac14}\vt, \qquad D\vt = \sqrt{2}x^{\frac14},
\end{equation}
and thus
\begin{equation}
J'(x,\vt)  = \frac{1}{2\sqrt{x}}\left(\vt j\big(\sqrt{x}\big) + \sqrt{2}\,x^{\frac14}\chi\big(\sqrt{x}\big)\right) \equiv \frac{1}{\sqrt 2}\mJ(x,\vt).
\end{equation}
In order for $\mJ(x,\vt)$ to define a consistent super-conformal theory on the supermanifold with local coordinates $(x,\vt),$ we need to require that the expansion of $\mJ$ into powers of $x$ consist of terms
with integer of half-integer powers. This implies that 
\begin{equation}
\chi(z) = \sum\limits_{m\in\mathbb Z}\chi_m\,z^{-m-\frac12}
\end{equation}
and consequently
\begin{equation}
\mJ(x,\vt) =  \frac{1}{\sqrt 2}\vt \sum\limits_{m\in\mathbb Z}\frac{a_m}{x^{\frac{m}{2}+1}} + \sum\limits_{m\in\mathbb Z}\frac{\chi_m}{x^{\frac{m}{2} + \frac12}}.
\end{equation}
There are four cases we shall discuss separately below.
\subsubsection{The NS-NS sector}
In the NS-NS case both components of $\mJ$ are invariant under $x \to \mathrm{e}^{2\pi i}x,$
\begin{equation}
\mJ(x,\vt) 
= 
\sum\limits_{m\in\mathbb Z}\frac{\chi_{2m+1}}{x^{m+1}} + \frac{\vt}{\sqrt 2} \sum\limits_{m\in\mathbb Z}\frac{a_{2m}}{x^{m+1}}
\equiv 
\sum\limits_{k\in\mathbb Z + \frac12}\frac{\psi_k}{x^{k+\frac12}} + \vt \sum\limits_{m\in\mathbb Z}\frac{\sa_m}{x^{m+1}} 
\equiv
\psi_{\NS}(x) + \vt \jmath_{\NS}(x)
\end{equation}
with
\begin{equation}
\label{eq:super_commutataors_NS_NS}
\sa_m = \frac{1}{\sqrt 2} a_{2m}, \quad \psi_k = \chi_{2k}, \qquad [\sa_m,\sa_n] = m\delta_{m+n,0}, \quad \{\psi_k,\psi_l\} = \delta_{k+l,0}.
\end{equation}
Notice that, due to a different normalization of the modes $\sa_m,$ the current $\jmath_{\NS}(x)$ used here is equal to the current defined in eq.\ (\ref{eq:bosonc_current_on_x_plane}) multiplied by $\sqrt{2}.$
Throughout the supersymmetric sections we use the normalisation
$[\sa_r,\sa_s]=r\delta_{r+s,0}$, and hence
$\sa_r=a_{2r}/\sqrt{2}$, for both integer and half-integer bosonic modes.
The purely bosonic subsections instead use $\sa_r=a_{2r}/2$ and
$[\sa_r,\sa_s]=(r/2)\delta_{r+s,0}$.  The two conventions differ only by
the overall factor $\sqrt{2}$ in the corresponding bosonic current.
With the standard definition of the normal ordering operation of the fermionic modes,
\begin{equation}
\label{eq:nomal_ordering_NS_fermions}
:\!\psi_k\psi_l\!: \; =
\left\{
\begin{array}{rl}
-\psi_l\psi_k & \mathrm{for}\ k>0, l < 0,
\\[2pt]
\psi_k\psi_l & \mathrm{otherwise},
\end{array}
\right.
\end{equation}
we have: 
\begin{equation}
\chi(y)\chi(x) =\; :\chi(y)\chi(x)\!: + \frac{1}{y-x}.
\end{equation}
Consequently
\begin{equation}
\label{eq:normal_ordering_JJ}
\mJ(y,\eta)\mJ(x,\vt) =\;:\!\mJ(y,\eta)\mJ(x,\vt)\!: + \frac{1}{y-x-\eta\vt}
\end{equation}
and
\begin{equation}
\label{eq:normal_ordering_DJJ}
\mDJ(y,\eta)\mJ(x,\vt) =\;:\!\mDJ(y,\eta)\mJ(x,\vt)\!: - \frac{\eta-\vt}{(y-x)^2},
\end{equation}
where
\begin{equation}
\mDJ(y,\eta) = \left(\frac{\partial}{\partial\eta} + \eta\frac{\partial}{\partial y}\right) \mJ(y,\eta) = \jmath_{\NS}(y) + \eta\partial\psi_{\NS}(y).
\end{equation}
For the future convenience we denote the modes of expansions of $\mJ(x,\vt)$ and $\mDJ(x,\vt)$ into the power series of $x$ as $\mJ_k(\vt)$ and $\mDJ_m(\vt),$ so that
\begin{equation}
\mJ(x,\vt) = \sum\limits_{k\in\mathbb Z+\frac12}\frac{\mJ_k(\vt)}{x^{k+\frac12}}, \qquad \mDJ(x,\vt) = \sum\limits_{m\in\mathbb Z}\frac{\mDJ_m(\vt)}{x^{m+1}}.
\end{equation}
Notice that
\begin{equation}
\mJ_k(\vt)\Big|_{\vt = 0} =\; \psi_k, \qquad \mDJ_m(\vt)\Big|_{\vt = 0} =\; \sa_m.
\end{equation}
From (\ref{eq:normal_ordering_JJ}) and (\ref{eq:normal_ordering_DJJ}) we get 
\begin{eqnarray}
\label{eq:commututors_JJ_DJJ_NSNS}
\nonumber
\{\mJ_k(\eta),\mJ_l(\vt)\} & = & \Res_{x=0}\Res_{y=x} \frac{x^{l-\frac12}y^{k-\frac12}}{y-x-\eta\theta} = 
\delta_{k+l,0}
+
\left(k-\frac12\right)\eta\vartheta
\delta_{k+l,1}
\\[-6pt]
\\[-6pt]
\nonumber
\left[\mDJ_m(\eta),\mJ_k(\vt)\right] & = & -(\eta-\vt)\Res_{x=0}\Res_{y=x} \frac{x^{k-\frac12} y^{m} }{(y-x)^2} =  -m(\eta-\vt)\delta_{m+k,\frac12},
\end{eqnarray}
while the OPE
\begin{equation}
\mDJ(y,\eta)\mDJ(x,\vt) \sim  \frac{1}{(y-x-\eta\vt)^2}
\end{equation}
obtained by applying $\frac{\partial}{\partial\vt} + \vt\frac{\partial}{\partial x}$ to (\ref{eq:normal_ordering_DJJ}), yields
\begin{equation}
\label{eq:commututors_DJDJ_NSNS}
\left[\mDJ_m(\eta),\mDJ_n(\vt)\right] = \Res_{x=0}\Res_{y=x} \frac{x^{n}y^{m}}{(y-x-\eta\theta)^2} = m\delta_{m+n,0} +  m(m-1)\eta\theta \delta_{m+n,1}.
\end{equation}

We now define the super-stress tensor
\begin{eqnarray}
\mT(x,\vt) & = & \lim\limits_{y\to x} \mDJ(y,\vt)\mJ(x,\vt) =\;:\!\mDJ(x,\vt)\mJ(x,\vt)\!:
\\[2pt]
\nonumber
& = & \jmath_{\NS}(x)\psi_{\NS}(x) + \vt\Big(\hskip -2pt :\!\jmath_\NS(x)\jmath_\NS(x)\!:\; + :\!\partial\psi_{\NS}(x)\psi_{\NS}(x)\!:\hskip -3pt\Big)
\end{eqnarray}
and compute its OPE with itself
\begin{equation}
\label{eq:NSNS_mTmT_OPE}
\mT(y,\eta)\mT(x,\vt)
\; \sim \;
\frac{1}{(y-x-\eta\vt)^3} 
+
\frac{3(\eta - \vt)}{(y-x)^2}\mT(x,\vt)
+
\frac{\mathcal{DT}(x,\vt)}{y-x-\eta\vt}
+
\frac{2(\eta-\theta)}{y-x}\frac{\partial\mT(x,\vt)}{\partial x}.
\end{equation}
If we expand
\begin{equation}
\mT(x,\vt) = \sum\limits_{k\in\mathbb Z+\frac12}\frac{\mT_k(\vt)}{x^{k+\frac32}}, \qquad  \mDT(x,\vt) = \sum\limits_{m\in\mathbb Z} \frac{\mDT_m(\vt)}{x^{m+2}},
\end{equation}
then the identity
\begin{equation}
\{\mT_k(\eta),\mT_l(\vt)\} = \Res_{x=0}\Res_{y=x}\, y^{k+\frac12} x^{l+\frac12 }\mT(y,\eta)\mT(x,\vt) 
\end{equation}
applied to (\ref{eq:NSNS_mTmT_OPE}) gives
\begin{eqnarray}
\label{eq:super_T_commutator}
\nonumber
&& \hskip -2cm
\{\mT_k(\eta),\mT_l(\vartheta)\}
=
\frac12
\left(k^2-\frac14\right)\delta_{k+l,0}
+
\frac12
\left(k^2-\frac14\right)\left(k-\frac32\right)
\eta\vartheta\,
\delta_{k+l-1,0}
\\[-6pt]
\\[-6pt]
\nonumber
&+ &
\left(k-2l-\frac12\right)(\eta-\vartheta)
\mT_{k+l-\frac12}(\vartheta)
+
\mDT_{k+l}(\vartheta)
+
\left(k+\frac12\right)
\eta\vartheta\,
\mDT_{k+l-1}(\vartheta).
\end{eqnarray}
Denoting
\begin{equation}
\mT_k(\vt)\Big|_{\vt = 0} = G_k, \qquad \mDT_m(\vt)\Big|_{\vt = 0} = 2L_m,
\end{equation}
one checks that eq.\ (\ref{eq:super_T_commutator}) is equivalent to
\begin{eqnarray}
\nonumber
\{G_k,G_l\} & = & 2L_{k+l} +  \frac12 \left(k^2-\frac14\right)\delta_{k+l,0},
\\
\left[L_m,G_k\right] & = & \left(\frac{m}{2}-k\right)G_{m+k},
\\
\nonumber
[L_m,L_n] & = & (m-n)L_{m+n} + \frac18 m\left(m^2-1\right)\delta_{m+n,0},
\label{eq:N_1_NS_NS_super}
\end{eqnarray}
where $k,l \in \mathbb{Z}+\frac12,\ m,n \in \mathbb{Z},$ which is the standard form of the NS super-algebra with the cental charge $c = \frac32.$ 

Let us now decompose the currents $\mJ$ and $\mDJ$ into terms singular and regular for $x\to 0,$
\begin{equation}
\mJ(x,\vt) = \mJ_>(x,\vt) + \mJ_<(x, \vt), \qquad \mJ_>(x,\vt) = \sum\limits_{k \geqslant \frac12} \frac{\mJ_k(\vt)}{x^{k+\frac12}}
\end{equation}
and
\begin{equation}
\mDJ(x,\vt) = \mDJ_\geqslant(x,\vt) + \mDJ_<(x, \vt), \qquad \mDJ_\geqslant(x,\vt) = \sum\limits_{m \geqslant 0} \frac{\mDJ_m(\vt)}{x^{m+1}}.
\end{equation}
Consequently
\begin{equation}
\mT(x,\vt) =  \mDJ_{\!\geqslant}(x,\vt)\mJ_>(x,\vt) + \mDJ_{\!<}(x,\vt)\mJ_>(x,\vt) + \mJ_<(x,\vt)\mDJ_{\!\geqslant}(x,\vt) +  \mDJ_{\!<}(x,\vt)\mJ_<(x,\vt)
\end{equation}
and the component of $\mT(x,\vt)$ containing terms singular for $x\to 0$ can be presented as
\begin{equation}
\mT^{\geqslant -\frac12}(x,\vt) 
= 
\mDJ_\geqslant(x,\vt)\mJ_>(x,\vt) + 
\Res\limits_{y=0}\frac{1}{x-y}\Big(\mDJ_<(y,\vt)\mJ_>(y,\vt) + \mJ_<(y,\vt)\mDJ_\geqslant(y,\vt)\Big).
\end{equation}

\subsubsection{The Ramond-Ramond sector}
In the R-R case both components of $\mJ$ change sign under $x \to \mathrm{e}^{2\pi i}x,$
\begin{equation}
\label{eq:super_RR_current_on_x_plane}
\mJ(x,\vt) 
= 
\sum\limits_{m\in\mathbb Z}\frac{\chi_{2m}}{x^{m+\frac12}} + \frac{\vt}{\sqrt 2} \sum\limits_{m\in\mathbb Z}\frac{a_{2m-1}}{x^{m+\frac12}}
\equiv 
\sum\limits_{m\in\mathbb Z}\frac{\psi_m}{x^{m+\frac12}} + \vt \hskip -3pt\sum\limits_{k\in\mathbb Z+\frac12}\frac{\sa_k}{x^{k+1}}
\equiv
\psi_{\R}(x) + \vt \jmath_{\R}(x)
\end{equation}
with
\begin{equation}
\sa_k = \frac{1}{\sqrt 2} a_{2k}, \quad \psi_k = \chi_{2k}, \qquad [\sa_k,\sa_l] = k\delta_{k+l,0}\, \quad \{\psi_m,\psi_n\} = \delta_{m+n,0}.
\end{equation}
If we define the normal ordering operation as
\begin{equation}
:\!\sa_k\sa_l\!:
=
\left\{
\begin{array}{rl}
\sa_l\sa_k & \mathrm{for}\; k > 0, l < 0,
\\[2pt]
\sa_k\sa_l & \mathrm{otherwise},
\end{array}
\right.
\hskip 1cm
:\!\psi_m\psi_n\!: \;=
\left\{
\begin{array}{rl}
0 &  \mathrm{for}\; m=n=0,
\\[2pt]
-\psi_n\psi_m & \mathrm{for}\; m \geqslant 0, n < 0,
\\[2pt]
\psi_m\psi_n & \mathrm{otherwise},
\end{array}
\right.
\end{equation}
then
\begin{equation}
\jmath_\R(y)\jmath_\R(x) = \; :\!\jmath_\R(y)\jmath_\R(x)\!: +  \frac{y+x}{2\sqrt{y}\sqrt{x}}\,\frac{1}{(y-x)^2}
= \frac{1}{(y-x)^2} + \ :\!\jmath_\R(y)\jmath_\R(x)\!: + \frac{1}{8x^2} + \mathcal{O}(1)
\end{equation}
and
\begin{equation}
\psi_\R(y)\psi_\R(x) =\;  :\!\psi_\R(y)\psi_\R(x)\!: + \frac{y+x}{2\sqrt{y}\sqrt{x}}\,\frac{1}{y-x}
=
\frac{1}{y-x} +\ :\!\psi_\R(y)\psi_\R(x)\!: +  \frac18 \frac{y-x}{x^2} + \mathcal{O}\left((y-x)^2\right).
\end{equation}
In effect, if we write down only the OPE terms singular in the limit $y\to x,$ then the results
\begin{equation}
\mJ(y,\eta)\mJ(x,\vt) \sim  \frac{1}{y-x-\eta\vt},
\qquad
\mDJ(y,\eta)\mJ(x,\vt) \sim - \frac{\eta-\vt}{(y-x)^2},
\end{equation}
and
\begin{equation}
\mDJ(y,\eta)\mDJ(x,\vt) \sim \frac{1}{(y-x-\eta\vt)^2},
\end{equation}
coincide with the OPE-s in the NS-NS sector. This suffices to conclude that also in the RR sector the modes $\mT_k(\vt)$ of super-stress tensor
\begin{equation}
\mT(x,\vt) = \lim\limits_{y\to x} \mDJ(y,\vt)\mJ(x,\vt), \qquad \mT_k(\vt) = \Res_{x=0}x^{k+\frac12}\mT(x,\vt), \quad k \in \mathbb{Z} + \frac12,
\end{equation}
satisfy the algebra (\ref{eq:super_T_commutator}). Moreover, even if the expansion of $\mJ(x,\vt)$ and $\mDJ(x,\vt)$ into the $x-$series contains half-integer rather than integer modes
\begin{equation}
\mJ(x,\vt) = \sum\limits_{m\in\mathbb Z}\frac{\mJ_m(\vt)}{x^{m+\frac12}}, \qquad \mDJ(x,\vt) = \sum\limits_{k\in\mathbb Z+\frac12}\frac{\mDJ_k(\vt)}{x^{k+1}},
\end{equation}
the algebra satisfied by $\mJ_m(\vt)$ and $\mDJ_k(\vt)$ is just an obvious modification of the algebra (\ref{eq:commututors_JJ_DJJ_NSNS}),  (\ref{eq:commututors_DJDJ_NSNS}). In particular
\begin{equation}
\label{eq:Jm_Jn_supercommutators}
\{\mJ_m(\eta),\mJ_n(\vt)\} = \Res_{x=0}\Res_{y=x}\frac{x^{n-\frac12}y^{m-\frac12}}{y-x-\eta\vt} = \delta_{m+n,0} + \left(m-\frac12\right)\eta\theta\,\delta_{m+n,1}.
\end{equation}
\subsubsection{The R-NS  and NS-R sectors}
For both 
\begin{equation}
\label{eq:Jcurrent_R-NS}
\mJ(x,\vt) 
=  
 \sum\limits_{k\in\mathbb Z+\frac12}\left(\vt\frac{\sa_k}{x^{k+1}}+\frac{\psi_k}{x^{k+\frac12}}\right)
\equiv
 \vt \jmath_{\R}(x) + \psi_{\NS}(x) 
\end{equation}
and
\begin{equation}
\label{eq:Jcurrent_NS-R}
\mJ(x,\vt) 
=  
 \sum\limits_{m\in\mathbb Z}\left(\vt\frac{\sa_m}{x^{m+1}}+\frac{\psi_m}{x^{m+\frac12}}\right)
\equiv
 \vt \jmath_{\NS}(x) + \psi_{\R}(x) 
\end{equation}
super-currents, where
\begin{equation}
 \left[\sa_s,\sa_t\right] = s\delta_{s+t,0}, \quad \{\psi_s,\psi_t\} = \delta_{s+t,0},
 \qquad s,t \in \mathbb{Z} + \frac12\;\; \mathrm{or}\;\; s,t \in \mathbb{Z}
\end{equation}
the $\mJ(y,\eta)\mJ(x,\vt)$ OPE is of the standard form. 
\begin{equation}
\mJ(y,\eta)\mJ(x,\vt) \sim \frac{1}{y-x-\eta\vt},
\end{equation}
Hence the super-stress tensor is still defined as
\begin{equation}
\mT(x,\vt) = \lim\limits_{y\to x} \mDJ(y,\vt)\mJ(x,\vt)
\end{equation}
and satisfies the OPE given in eq. (\ref{eq:NSNS_mTmT_OPE}). However, if we decompose
\begin{equation}
\mT(x,\vt) = 2\vt T(x) + G(x),
\end{equation}
where
\begin{equation}
G(x) = \psi_{\NS/\R}(x)\jmath_{\R/\NS}(x),
\qquad
2T(x) = \lim\limits_{y\to x}\left(\jmath_{\R/\NS}(y)\jmath_{\R/\NS}(x) 
+ 
\partial\psi_{\NS/\R}(y)\psi_{\NS/\R}(x)\right),
\end{equation}
then, even if the expansion of $T$ into $x$ series contains integer powers, the power series expansion of $G$ contains half-integer modes
\begin{eqnarray}
T(x) = \sum\limits_{m\in\mathbb Z}\frac{L_m}{x^{m+2}}, \qquad G(x) = \sum\limits_{m\in\mathbb Z}\frac{G_m}{x^{m+\frac32}}.
\end{eqnarray}
This expansion along with the OPE  (\ref{eq:NSNS_mTmT_OPE}) leads to the (super)algebra of the modes
\begin{eqnarray}
\label{eq:Ramond_superalgebra}
\nonumber
[L_m,L_n] & = & (m-n)L_{m+n} + \frac{1}{8}m\left(m^2-1\right)\delta_{m+n,0},
\\[0pt]
[L_m,G_n] & = & \frac{m-2n}{2}G_{m+n}, 
\\
\nonumber
\{G_m,G_n\} & = & 2L_{m+n} + \frac{1}{2}\left(m^2-\tfrac14\right)\delta_{m+n,0},
\end{eqnarray}
being the Ramond algebra with the central charge $c = \frac32.$

\section{Miwa transform and Kontsevich's operators}
\label{sec: Section 3}

Hawing constructed the super-formalism on the bosonic surface, we now want to study, how the theory reduce or simplify when we consider Miwa variables \cite{Miwa82}, which one can understand as an odd traces \cite{Kontsevich92} used by Kontsevich in order to proof the Witten's conjecture. Before we move into the super-symmetric sector, we want to show, how such transformations, defined in a pure bosonic setting into NS and R sectors, simplify the energy-momentum current and positive part of a current $j(z)$, which in future would be a motivation on introducing super-Miwa transformation.

\setcounter{equation}{0}
\subsection{Bosonic case}
We shall start by discussing the NS sector.
Commutation relations satisfied by the modes of the field $\jmath_\NS(x),$ given in eq.\ (\ref{eq:bosonc_current_on_x_plane}), can be represented on the space of function of commuting variables 
$g_m, m \geqslant 0,$ as
\begin{equation}
\label{eq:first_repesentation_boson}
\sa_m = \frac{\partial}{\partial g_m},\; m \geqslant 0, \qquad \sa_{-m} = \frac{m}{2}g_m,\; m \geqslant 1.
\end{equation}

CFT-inspired way of formulating Miwa transformation consist of replacing $g_m$ with symmetric function of variables $l_i$ in such a way, that
\begin{equation}
\label{eq:Miwa_definition_boson_NS}
\jmath_{\NS,\geqslant}(x)\Big|_{x=l_i} = -\frac{\partial}{\partial l_i}.
\end{equation}
For $\jmath_{\NS,\geqslant}(l_i) = \sum\limits_{n\geqslant 0} l_i^{-n-1}\sa_n,$  using (\ref{eq:first_repesentation_boson}), (\ref{eq:Miwa_definition_boson_NS}) 
and the chain rule for change of variables
\begin{equation}
\label{eq:NS_Miwa_derivation}
-\frac{\partial}{\partial l_i} = \sum\limits_{m\geqslant 0}l_i^{-m-1}\frac{\partial}{\partial g_m} 
= 
\frac{\partial\log l_i}{\partial l_i}\frac{\partial}{\partial g_0}
-  \sum\limits_{m\geqslant 0}\frac{1}{m}\frac{\partial l_i^{-m}}{\partial l_i}\frac{\partial}{\partial g_m}
=
-\sum\limits_{m\geqslant 0}\frac{\partial g_m}{\partial l_i}\frac{\partial}{\partial g_m}
\end{equation}
we get an explicit form of  Miwa transformation in NS sector:
\begin{equation}
g_0 = -\sum\limits_j\log l_j, \qquad g_m = \frac{1}{m}\sum\limits_j l_j^{-m},\quad m \geqslant 1.
\label{eq:NS_Miwa_transformation}
\end{equation}
The creation (negative mode) part of free boson current assumes in the Miwa variables a simple form 
\begin{equation}
\jmath_{\NS,<}(x) = \sum\limits_{m\geqslant 1}\frac{m}{2}g_m\, x^{m-1} = \frac12\sum\limits_j\sum\limits_{m\geqslant 1} l_j^{-m}x^{m-1} = \frac12\sum\limits_j \frac{1}{l_j-x}.
\end{equation}
Next
\begin{equation}
\frac{\partial^2}{\partial l_i^2}
= 
-\frac{\partial}{\partial l_i}\sum\limits_{m\geqslant 0}\frac{1}{l_i^{m+1}}\frac{\partial}{\partial g_m}
=
\sum\limits_{n\geqslant 0}\frac{1}{l_i^{n+1}} \frac{\partial}{\partial g_n} \sum\limits_{m\geqslant 0}\frac{1}{l_i^{m+1}}\frac{\partial}{\partial g_m}
-
\sum\limits_{m\geqslant 0}\left[\frac{\partial}{\partial l_i},\frac{1}{l_i^{m+1}}\right]\frac{\partial}{\partial g_m},
\end{equation}
and consequently
\begin{equation}
\Big(\jmath_{\NS,\geqslant}(l_i)\Big)^2 = \frac{\partial^2}{\partial l_i^2} + \partial\jmath_{\NS,\geqslant}(x)\Big|_{x=l_i}.
\end{equation}
Since
\begin{eqnarray}
&&
\hskip -.5cm
2\Res_{y=0}\frac{\jmath_{\NS,<}(y)\jmath_{\NS,\geqslant}(y)}{l_i-y}
=
\Res_{y=0}\sum\limits_j\frac{\jmath_{\NS,\geqslant}(y)}{(y-l_i)(y-l_j)}
=
-\sum\limits_j\left(\Res_{x=l_i}+\Res_{y=l_j}\right)\frac{\jmath_{\NS,\geqslant}(y)}{(y-l_i)(y-l_j)}
\\[2pt]
\nonumber
& = & 
- \sum\limits_{j;\,j\neq i}\left(\frac{\jmath_{\NS,\geqslant}(l_i)}{l_i-l_j} + \frac{\jmath_{\NS,\geqslant}(l_j)}{l_j-l_i}\right)
-\Res_{y=l_i}\frac{\jmath_{\NS,\geqslant}(y)}{(y-l_i)^2}
=
\sum\limits_{j;\,j\neq i} \frac{1}{l_i-l_j}\left(\frac{\partial}{\partial l_i} - \frac{\partial}{\partial l_j}\right)-\partial\jmath_{\NS,\geqslant}(y)\Big|_{y=l_i},
\end{eqnarray}
where we used the fact that $\frac{\jmath_{\NS,<}(y)\jmath_{\NS,\geqslant}(y)}{l_i-y}$ is an analytic, regular at
infinity function of $y$ so that its residues at the finite part of the complex $y$ plane add up to zero,
we arrive at the formula
\begin{equation}
\label{eq:bosonic_singular_TNS}
\sT_{\NS}^{\geqslant -1}(l_i) 
= 
\Big(\jmath_{\NS,\geqslant}(l_i)\Big)^2 + 2\Res_{y=0}\frac{\jmath_{\NS,<}(y)\jmath_{\NS,\geqslant}(y)}{l_i-y} 
=
\frac{\partial^2}{\partial l_i^2} + \sum\limits_{j;\,j\neq i} \frac{1}{l_i-l_j}\left(\frac{\partial}{\partial l_i}- \frac{\partial}{\partial l_j}\right).
\end{equation}

Computations in the R sector are similar. Representing
\begin{equation}
\sa_k = \frac{\partial}{\partial g_k},\qquad \sa_{-k} = \frac{k}{2}g_k, \qquad k \in \mathbb{N}+\frac12
\end{equation}
and requiring that
\begin{equation}
\label{eq:Miwa_definition_boson_R}
\jmath_{\R,>}(l_i) \equiv \jmath_{\R,>}(x)\Big|_{x=l_i} = -\frac{\partial}{\partial l_i},
\end{equation}
we conclude that $g_k = \frac{1}{k}\sum\limits_j l_j^{-k}.$ In effect
\begin{equation}
\jmath_{\R,<}(x) = \frac12\sum\limits_j \sum\limits_{k\geqslant \frac12}l_j^{-k} x^{k-\frac12} = \frac12\sum\limits_j\frac{\sqrt{l_j}}{\sqrt{x}}\frac{1}{l_j-x}.
\end{equation}
Since
\begin{eqnarray}
&&
\hskip -.5cm
2\,\Res_{y=0}\frac{\jmath_{\R,<}(y)}{l_i-y}\jmath_{\R,>}(y)
=
\sum\limits_j\Res_{y=0}\frac{\sqrt{l_j}}{(l_i-y)(l_j-y)}\frac{\jmath_{\R,>}(y)}{\sqrt{y}}
\\[4pt]
\nonumber
& = & 
- \sum\limits_j\left(\Res_{y=l_i}+\Res_{y=l_j}\right)\frac{\sqrt{l_j}}{(l_i-y)(l_j-y)}\frac{\jmath_{\R,>}(y)}{\sqrt{y}}
=
\sum\limits_{j,\,j\neq i}\frac{1}{l_i-l_j}\left(\frac{\sqrt{l_j}}{\sqrt{l_i}}\frac{\partial}{\partial l_i} - \frac{\partial}{\partial l_j}\right)
-
\frac{1}{2l_i}\frac{\partial}{\partial l_i} - \partial\jmath_{\R,>}(l_i),
\end{eqnarray}
we get
\begin{eqnarray}
\sT_\R^{\geqslant -1}(l_i)
& = &
\frac{\partial^2}{\partial l_i^2} 
- 
\frac{1}{2l_i}\frac{\partial}{\partial l_i} 
+
\sum\limits_{j,\,j\neq i}\frac{\sqrt{l_j}}{l_i-l_j}\left(\frac{1}{\sqrt{l_i}}\frac{\partial}{\partial l_i} - \frac{1}{\sqrt{l_j}}\frac{\partial}{\partial l_j}\right)
+
\frac{1}{4l_i}\Big(\sum\limits_j\frac{1}{\sqrt{l_j}}\Big)^2 
+
\frac{1}{16 l_i^2}
\\[4pt]
\nonumber
& = &
\sT_{\NS}^{\geqslant -1}(l_i) 
-
\frac{1}{\sqrt{l_i}}\sum\limits_j\frac{1}{\sqrt{l_i}+\sqrt{l_j}}\frac{\partial}{\partial l_i}
+
\frac{1}{4l_i}\Big(\sum\limits_j\frac{1}{\sqrt{l_j}}\Big)^2 
+
\frac{1}{16 l_i^2}.
\end{eqnarray}
In order to get rid of the square roots we may change variables and define $l_i = \lambda_i^2.$ In effect,
we will reproduce the result of \cite{IZ92}.

\subsection{The super-conformal case}
\subsubsection{The NS-NS sector}

We represent the $\mJ_{-l}(0), k \geqslant \frac12$ mode as multiplication by an odd variable $q_l$ and the $\mDJ_{-m}(0)$ mode 
(with $m \geqslant 1$) as a multiplication by an even variable $m g_m,$ so that
\begin{equation}
\mJ_{-l}(\vt) = \mJ_{-l}(0) + \vt \mDJ_{-l-\frac12}(0) = q_l + \vt\left(l+\frac12\right)g_{l+\frac12},
\end{equation}
and thus, using eq.\ (\ref{eq:commututors_JJ_DJJ_NSNS}),
\begin{equation}
\mJ_k(\eta) = \frac{\partial}{\partial q_k} + \eta\frac{\partial}{\partial g_{k-\frac12}}, \qquad k \geqslant\frac12.
\end{equation}
We shall define a SUSY analog of the Miwa transformation by introducing, along with a set of even variables $l_i,$  an accompanying set of odd variables $\theta_i$ and 
requiring, that for all values of the index $i:$
\begin{eqnarray}
\mJ_>(l_i,\theta_i) \stackrel{!}{=} - \frac{\partial}{\partial\theta_i}- \theta_i \frac{\partial}{\partial l_i} \equiv -D_i.
\end{eqnarray}
Performing a computation analogous to  (\ref{eq:NS_Miwa_derivation}) we get
\begin{equation}
g_0 = -\sum\limits_j\log l_j, \quad g_m = \frac{1}{m}\sum\limits_j l_j^{-m},\; m \geqslant 1, \quad q_k = -\sum\limits_j \theta_j l_j^{-k-\frac12}, \; k \geqslant \frac12.
\end{equation}
In effect
\begin{eqnarray}
\nonumber
\mDJ_\geqslant(l_i,\theta_i) & = & D_i\mJ_>(l_i,\theta_i) = -D_i^2 = -\frac{\partial}{\partial l_i},
\\[-6pt]
\\[-6pt]
\nonumber
\mJ_<(y,\theta) 
& = & 
\sum\limits_{k\geqslant \frac12}y^{k-\frac12}\mJ_{-k}(\theta)
=
\sum\limits_j(\theta - \theta_j)\sum\limits_{k\geqslant \frac12} y^{k-\frac12}l_j^{-k-\frac12} = -\sum\limits_{j}\frac{\theta - \theta_j}{y-l_j}
\end{eqnarray}
and thus
\begin{equation}
\qquad
\mDJ_<(y,\theta)  =  \left(\frac{\partial}{\partial\theta}+ \theta \frac{\partial}{\partial y}\right)\mJ_<(y,\theta)  =  -\sum\limits_j\frac{1}{y-l_j-\theta\theta_j}.
\end{equation}

We have
\begin{eqnarray}
\nonumber
&&
\hskip -1cm 
\Res\limits_{y=0}\frac{\mDJ_<(y,\theta_i)}{l_i-y}\mJ_>(y,\theta_i)
=
-\hskip -1pt \sum\limits_j\left(\Res_{y=l_i} + \hskip -5pt \Res_{y=l_j+\theta_i\theta_j}\right)\hskip -2pt\frac{\mJ_>(y,\theta_i)}{(y-l_i)(y-l_j-\theta_i\theta_j)}
\\[2pt]
\nonumber
& = &
\hskip -5pt
\sum\limits_{j;\,j\neq i}\hskip -3pt\frac{-\mJ_>(l_i,\theta_i) + \mJ_>(l_j+\theta_i\theta_j,\theta_i)}{l_i-l_j-\theta_i\theta_j} 
 - \partial\mJ_>(l_i,\theta_i)
\\[2pt]
& = &
\sum\limits_{j;\,j\neq i}\frac{-\mJ_>(l_i,\theta_i) + \mJ(l_j,\theta_j) + (\theta_i-\theta_j)\mDJ_{\geq}(l_j,\theta_j)}{l_i-l_j-\theta_i\theta_j} 
 - \partial\mJ_>(l_i,\theta_i) 
 \\[2pt]
\nonumber
& = &
\sum\limits_{j;\,j\neq i}\left(\frac{D_i-D_j}{l_i-l_j-\theta_i\theta_j}-\frac{\theta_i-\theta_j}{l_i-l_j}\frac{\partial}{\partial l_j} \right)
 - \partial\mJ_>(l_i,\theta_i) 
\end{eqnarray}
and similarly
\begin{eqnarray}
&& \hskip -2cm
\Res\limits_{y=0}\frac{\mJ_<(y,\theta_i)}{l_i-y}\mDJ_>(y,\theta_i)
=
\sum\limits_j\Res_{y=0}\frac{(\theta_i-\theta_j)\mDJ_>(y,\theta_i)}{(y-l_i)(y-l_j)}
\\[2pt]
\nonumber
& = &
-\sum\limits_{j;\,j\neq i}\left(\Res_{y=l_i}+\Res_{y=l_j}\right)\frac{(\theta_i-\theta_j)\mDJ_>(y,\theta_i)}{(y-l_i)(y-l_j)}
=
\sum\limits_{j;\,j\neq i}\frac{\theta_i-\theta_j}{l_i-l_j}\left(\frac{\partial}{\partial l_i} - \frac{\partial}{\partial l_j}\right).
\end{eqnarray}
Finally
\begin{equation}
-\frac{\partial}{\partial l_i}\left(-\frac{\partial}{\partial\theta_i} - \theta_i\frac{\partial}{\partial l_i}\right)
=
\mDJ_>(l_i,\theta_i)\mJ_>(l_i,\theta_i) - \partial\mJ_>(l_i,\theta_i)
\end{equation}
so that
\begin{equation}
\mDJ_>(l_i,\theta_i)\mJ_>(l_i,\theta_i) = \frac{\partial}{\partial l_i}D_i
+
\partial\mJ_>(l_i,\theta_i)
=
D_i^3 + \partial\mJ_>(l_i,\theta_i).
\end{equation}
Collecting all the terms above we get
\begin{eqnarray}
&& \hskip -1.5cm
\mT^{\geqslant -\frac12}_{\NS\NS}(l_i,\theta_i)
=
D_i^3
+
\sum\limits_{j;\,j\neq i}\left(
\frac{\theta_i-\theta_j}{l_i-l_j} \left(\frac{\partial}{\partial l_i} - 2\frac{\partial}{\partial l_j}\right)
+
\frac{D_i - D_j}{l_i-l_j-\theta_i\theta_j}
\right)
\\[4pt]
\nonumber
& = &
\theta_i T_\NS(l_i)
+
\frac{\partial}{\partial l_i}\frac{\partial}{\partial\theta_i}
+
\sum\limits_{j;\,j\neq i}\frac{1}{l_i-l_j-\theta_i\theta_j}
\left(\frac{\partial}{\partial\theta_i} - \frac{\partial}{\partial\theta_j} +
(\theta_i-\theta_j)\left(\frac{\partial}{\partial l_i} - \frac{\partial}{\partial l_j}\right)
\right).
\end{eqnarray}

\subsubsection{The R-R sector}
We represent modes $\mJ_m(\vt)$ satisfying the algebra (\ref{eq:Jm_Jn_supercommutators}) as operators acting on space of functions of
independent, even variables $g_k,\ k \in \mathbb{Z}_{\geqslant 0} + \frac12$ and odd variables $q_m,\ m\in\mathbb{Z}_{\geqslant 0}$ as
\begin{equation}
\mJ_{-m}(\vt) = q_m + \left(m+\frac12\right)\vt g_{m+\frac12},\quad \mJ_m(\eta) = \frac{\partial}{\partial q_m} + \eta \frac{\partial}{\partial g_{m-\frac12}}, \quad m \geqslant 1,
\end{equation}
and
\begin{equation}
\mJ_0(\vt) = \frac{q_0}{2}  + \frac{\vt}{2} g_{\frac12}+ \frac{\partial}{\partial q_0}.
\end{equation}
Since the zero mode of the super-current contains operators of both multiplication and differentiation, we split it and define
\begin{equation}
\mJ(x,\vt) = \mJ_>(x,\vt) + \mJ_<(x,\vt),
\end{equation}
where
\begin{equation}
\mJ_>(x,\vt) = \frac{1}{\sqrt{x}}\frac{\partial}{\partial q_0} + \sum\limits_{m \geqslant 1}\frac{\mJ_m(\vt)}{x^{m+\frac12}},
\qquad
\mJ_<(x,\vt) = \frac{1}{2\sqrt{x}}\left(\vt g_{\frac12} + q_0\right) + \sum\limits_{m \geqslant 1}\mJ_{-m}(\vt)x^{m-\frac12}.
\end{equation}
Since
\begin{equation}
\left\{\mJ_>(y,\eta),\mJ_<(x,\vt)\right\} = \frac{y+x}{2\sqrt{y}\sqrt{x}}\frac{1}{y-x-\eta\vt}
\end{equation}
and in effect
\begin{eqnarray}
\nonumber
\left[\mDJ_>(y,\eta),\mJ_<(x,\vt)\right] & = & \frac{\eta}{4\sqrt{y^3}\sqrt{x}} - \frac{(\eta-\vt)(x+y)}{2\sqrt{x}\sqrt{y}(y-x)^2},
\\[-6pt]
\\[-6pt]
\nonumber
\left[\mJ_>(y,\eta),\mDJ_<(x,\vt)\right] & = & \frac{\vt}{4\sqrt{y}\sqrt{x^3}} + \frac{(\eta-\vt)(x+y)}{2\sqrt{x}\sqrt{y}(y-x)^2},
\end{eqnarray}
we see that the super-stress tensor can be written as
\begin{eqnarray}
\nonumber
\mT(x,\vt)
& = &
\mDJ_>(x,\vt)\mJ_>(x,\vt) + \mDJ_<(x,\vt)\mJ_>(x,\vt) + \mJ_<(x,\vt)\mDJ_>(x,\vt)
\\[4pt]
& + &
\mDJ_<(x,\vt)\mJ_<(x,\vt) + \frac{\vt}{4x^2}
\end{eqnarray}
Separating terms singular for $x\to 0$ we thus get
\begin{eqnarray}
\nonumber
\mT^{\geqslant-\frac12}(x,\vt)
& = &
\mDJ_>(x,\vt)\mJ_>(x,\vt) + \Res_{y=0}\frac{1}{x-y}\Big(\mDJ_<(y,\vt)\mJ_>(y,\vt) + \mJ_<(y,\vt)\mDJ_>(y,\vt)\Big)
\\[4pt]
& + &
\frac{\vt}{4x^2} + \frac{\vt}{4x}\left(g_{\frac12}^2 + 2q_1q_0\right) + \frac{1}{4x}q_0g_{\frac12}.
\end{eqnarray}

We now introduce, as in the NS-NS case, a set of even variables $l_i$ and odd variables $\theta_i,$ 
\begin{equation}
q_m = -\sum\limits_j \theta_j l_j^{-m-\frac12}, \qquad g_{m+\frac12} = \frac{1}{m+\frac12}\sum\limits_j l_j^{-m-\frac12}
\end{equation}
In effect
\begin{eqnarray}
\mJ_>(l_i,\theta_i) & = &\sum\limits_{m\geqslant 0}l_i^{-m-\frac12} \left(\frac{\partial}{\partial q_m} + \frac{\theta_i}{l_i} \frac{\partial}{\partial g_{m+\frac12}}\right)
\\[4pt]
\nonumber
& = &
-\sum\limits_{m\geqslant 0}\left(D_i q_m \frac{\partial}{\partial q_m} + D_i g_{m+\frac12} \frac{\partial}{\partial g_{m+\frac12}}\right)
= -D_i
\equiv
-\left(\frac{\partial}{\partial \theta_i} + \theta_i \frac{\partial}{\partial l_i}\right),
\end{eqnarray}
what implies that 
\begin{equation}
\mDJ_>(l_i,\theta_i) = \left(\frac{\partial}{\partial\vt} + \vt\frac{\partial}{\partial x}\right) \mJ_>(l_i,\theta_i)\Big|_{\stackrel{\vt = \theta_i}{x=l_i}}
=
-\frac{\partial}{\partial l_i}.
\end{equation}
We also have
\begin{eqnarray}
\mJ_<(x,\vt) 
& = &
\sum\limits_{m\geqslant 1}\mJ_{-m}(\vt)x^{m-\frac12} + \frac{1}{2\sqrt{x}}\left(q_0 + \vt g_{\frac12}\right)
\\[4pt]
\nonumber
& = &
\sum\limits_{m\geqslant 0}x^{m-\frac12}\left(q_m + \left(m+\frac12\right)\vt g_{m+\frac12}\right) -\frac{q_0}{2\sqrt{x}}
=
-\sum\limits_{j}\frac{\sqrt{l_j}}{\sqrt{x}}\frac{\vt-\theta_j}{x-l_j} -\frac{q_0}{2\sqrt{x}}
\end{eqnarray}
and
\begin{equation}
\mDJ_<(x,\vt) = -\sum\limits_j\frac{\sqrt{l_j}}{\sqrt{x-\vt\theta_j}}\frac{1}{x-l_j-\vt\theta_j} 
+
\frac{\vt q_0}{4\sqrt{x^3}}.
\end{equation}
It is not difficult, even if a bit tedious, to check the identities
\begin{eqnarray}
&&
\hskip -1cm
\Res_{y=0}\frac{\mJ_<(y,\theta_i)}{l_i-y}\mDJ_>(y,\theta_i)
=
\sum\limits_{j;j\neq i}\frac{(\theta_i-\theta_j)\sqrt{l_j}}{l_i-l_j}\left(\frac{1}{\sqrt{l_i}}\frac{\partial}{\partial l_i} - \frac{1}{\sqrt{l_j}}\frac{\partial}{\partial l_j}\right)
+
\frac{q_0}{2\sqrt{l_i}}\frac{\partial}{\partial l_i}
\\[4pt]
\nonumber
& = &
\sum\limits_{j;j\neq i}\frac{\theta_i-\theta_j}{l_i-l_j}
\left(\frac{\partial}{\partial l_i} -\frac{\partial}{\partial l_j}\right)
-
\frac{1}{\sqrt{l_i}}
\left(\sum\limits_{j;j\neq i}\frac{\theta_i-\theta_j}{\sqrt{l_i}+\sqrt{l_j}}
-
\frac{q_0}{2}\right)\frac{\partial}{\partial l_i}
\end{eqnarray}
and
\begin{eqnarray}
\nonumber
&&\hskip -2cm
\Res_{y=0}\frac{\mDJ_<(y,\theta_i)}{l_i-y}\mJ_>(y,\theta_i)
=
\\[6pt]
& = &
\sum\limits_{j;\,j\neq i}\frac{D_i-D_j}{l_i-l_j-\theta_i\theta_j}
-
\sum\limits_{j;\,j\neq i}\frac{\theta_i-\theta_j}{l_i-l_j}\frac{\partial}{\partial l_j}
-
\frac{1}{\sqrt{l_i}}\sum\limits_{j;\,j\neq i}\frac{\sqrt{l_i}-\sqrt{l_j}}{l_i-l_j-\theta_i\theta_j}D_i
\\[2pt]
\nonumber
& - &
\frac{D_i}{2l_i}
+
\frac{1}{2\sqrt{l_i^3}}\left(\frac{q_0}{2} - \sum\limits_{j;\,j\neq i}\frac{\theta_j\sqrt{l_j}}{l_i-l_j}\right)\theta_i\frac{\partial}{\partial\theta_i}
- 
\partial\mJ_>(l_i,\theta_i).
\end{eqnarray}
Since
\begin{equation}
\mDJ_>(l_i,\theta_i)\mJ_>(l_i,\theta_i) = D_i^3 + \partial\mJ_>(l_i,\theta_i)
\end{equation}
we arrive at the formula
\begin{eqnarray}
\label{eq:final_expression_for_TRR}
\nonumber
\mT^{\geqslant-\frac12}_{\R\R}(l_i,\theta_i)
& = &
D_i^3
+
\sum\limits_{j;\,j\neq i}\left(
\frac{\theta_i-\theta_j}{l_i-l_j} \left(\frac{\partial}{\partial l_i} - 2\frac{\partial}{\partial l_j}\right)
+
\frac{D_i - D_j}{l_i-l_j-\theta_i\theta_j}
\right)
+\delta \mT^{\geqslant-\frac12}(l_i,\theta_i)
\\[6pt]
& \equiv &
\mT^{\geqslant-\frac12}_{\NS\NS}(l_i,\theta_i)
+\delta \mT^{\geqslant-\frac12}(l_i,\theta_i)
\end{eqnarray}
where
\begin{eqnarray}
\nonumber
&&
\hskip -1cm
\delta \mT^{\geqslant-\frac12}(l_i,\theta_i) 
=
-
\left(\frac{1}{2l_i}
+
\sum\limits_{j;\,j\neq i}\frac{1}{\sqrt{l_i-\theta_i\theta_j}}\frac{\sqrt{l_i-\theta_i\theta_j}-\sqrt{l_j}}{l_i-l_j-\theta_i\theta_j}
\right)D_i
-
\frac{1}{\sqrt{l_i}}\sum\limits_{j;\,j\neq i}\frac{\theta_i - \theta_j}{\sqrt{l_i}+\sqrt{l_j}}\frac{\partial}{\partial l_i}
\\[4pt]
& + &
\frac{q_0}{2\sqrt{l_i}}\left(\frac{\partial}{\partial l_i} + \frac{\theta_i}{2l_i}\frac{\partial}{\partial\theta_i}\right)
+
\frac{\theta_i}{4l_i^2}+ \frac{\theta_i}{4l_i}\left(g_{\frac12}^2 + 2q_1q_0\right) + \frac{1}{4l_i}q_0g_{\frac12}.
\end{eqnarray}

\subsubsection{The R-NS sector}
The maximal, non-anomalous, containing positive modes, sub-algebra  of the Ramond algebra (\ref{eq:Ramond_superalgebra}) is spanned by $\{G_{\geqslant 0}, L_{\geqslant 1}\}\cup\{L_0-\tfrac{1}{16}\}.$ The strongest condition we can impose on the partition function thus reads
\begin{equation}
\left(\mT^{\geqslant 0}(x,\vt) -\frac{\vt}{8}\frac{1}{x^2} \right)Z
\equiv
\left(2\vt T_{\geqslant 0}(x) +G_{\geqslant 0}(x) -\frac{\vt}{8}\frac{1}{x^2} \right)Z = 0,
\end{equation}
where
\begin{equation}
G_{\geqslant 0}(x) = \sum\limits_{m\geqslant 0}\frac{G_m}{x^{m+\frac32}}, \quad T_\geqslant(z) = \sum\limits_{m\geqslant 0}\frac{L_m}{z^{m+2}}.
\end{equation}
We denote
\begin{equation}
\jmath_\R(x)= \jmath_<(x) + \jmath_>(x),
\qquad
\jmath_<(x)  = \sum\limits_{k\geqslant\frac12}\sa_{-k}x^{k-1},
\quad
\jmath_>(x)  = \sum\limits_{k\geqslant\frac12}\sa_{k}x^{-k-1},
\end{equation}
and
\begin{equation}
\psi_{\NS}(x) = \psi_<(x) + \psi_>(x),
\qquad
\psi_<(x) = \sum\limits_{k\geqslant\frac12}\psi_{-k}x^{k-\frac12},
\quad
\psi_>(x)  = \sum\limits_{k\geqslant\frac12}\psi_{k}x^{-k-\frac12}.
\end{equation}
Since
\begin{equation}
\left[\jmath_>(y),\jmath_<(x)\right] = \frac{y+x}{2\sqrt{y}\sqrt{x}}\frac{1}{(y-x)^2},
\qquad
\left\{\partial\psi_>(y),\psi_<(x)\right\} = - \frac{1}{(y-x)^2},
\end{equation}
while $\jmath$ and $\psi$ commute, we obtain 
\begin{eqnarray}
\nonumber
2T(x)
& = &
\jmath_<(x)\jmath_<(x) + 2\jmath_<(x)\jmath_>(x) + \jmath_>(x)\jmath_>(x) 
\\[4pt]
\nonumber
& + &
\partial\psi_<(x)\psi_<(x)+\partial\psi_<(x)\psi_>(x) -\psi_<(x)\partial\psi_>(x)+\partial\psi_>(x)\psi_>(x)+ \frac{1}{8x^2},
\\[4pt]
\nonumber
G(x) & = &  \psi_<(x)\jmath_<(x)+\psi_<(x)\jmath_>(x)+\jmath_<(x)\psi_>(x)+\psi_>(x)\jmath_>(x).
\end{eqnarray}
Since none of the terms 
\begin{equation}
\jmath_<(x)\jmath_<(x) \qquad \partial\psi_<(z)\psi_<(z) \qquad \mathrm{and} \qquad \psi_<(z)\jmath_<(x)
\end{equation}
contribute to $\mT^{\geqslant 0}_{\R\NS}(x,\vt),$ we get
\begin{eqnarray}
\label{eq:T_R_NS}
\nonumber
\mT^{\geqslant 0}_{\R\NS}(l_i,\theta_i) - \frac{\theta_i}{8 l_i^2}
& = &
\left(\jmath_>(x) + \theta_i\partial\psi_>(x)\right)\left(\theta_i\jmath_>(x) + \psi_>(x)\right)\Big|_{x=l_i}
\\[4pt]
& + &
\Res_{y=0}\frac{\sqrt{y}}{\sqrt{l_i}(l_i-y)}\Big(\psi_<(y)\jmath_>(y)+\jmath_<(y)\psi_>(y)\Big)
\\[4pt]
\nonumber
& + &
\theta_i \Res_{y=0}\frac{y}{{l_i}(l_i-y)}\Big(2\jmath_<(y)\jmath_>(y)+\partial\psi_<(y)\psi_>(y) -\psi_<(y)\partial\psi_>(y) \Big),
\end{eqnarray}
where the form of the $\Res$ operator in the second line follows from the fact that we are collecting singular terms in half-integer powers of $l_i$ starting form $l_i^{-\frac32}$ 
while the form of the $\Res$ operator in the third line follows from the fact that we are collecting there singular terms in integer powers of $l_i$ starting form $l_i^{-2}.$

The super-Miwa transformation,
\begin{equation}
\sa_{-k} = kg_{k} = \sum\limits_j\frac{1}{l_j^k}, \quad \sa_k = \frac{\partial}{\partial g_k},
\qquad 
\psi_{-k} = q_k = -\sum\limits_{j}\frac{\theta_j}{l_j^{k+\frac12}}, \quad \psi_k=\frac{\partial}{\partial q_k},
\end{equation}
now gives
\begin{equation}
\psi_>(x)\Big|_{x=l_i}
=
-\frac{\partial}{\partial\theta_i},
\qquad
\left(\jmath_>(x) + \theta_i\partial\psi_>(x)\right)\Big|_{x=l_i}
=
-\frac{\partial}{\partial l_i},
\end{equation}
and thus
\begin{equation}
\left(\jmath_>(x) + \vt\partial\psi_>(x)\right)\left(\vt\jmath_>(x) + \psi_>(x)\right)\Big|_{x=l_i}
=
\frac{\partial}{\partial l_i}\left(\frac{\partial}{\partial \theta_i} + \theta_i\frac{\partial}{\partial l_i}\right) + \partial\big(\theta_i\jmath_>(x) + \psi_>(x)\big)\Big|_{x=l_i}.
\end{equation}

To proceed let us rewrite the sum of the second and the third line of (\ref{eq:T_R_NS}) as
\begin{eqnarray}
\label{eq:T_R_NS_new}
&&
\hskip -2cm
\Res\limits_{y=0}\frac{1}{l_i-y}\left(\frac{\sqrt{y}}{\sqrt{l_i}}\psi_<(y)+ \frac{y}{l_i}\theta_i \jmath_<(y)\right)
\left(\jmath_>(y) + \frac{\sqrt{y}}{\sqrt{l_i}}\,\theta_i\partial\psi_>(y)\right)
\\[4pt]
\nonumber
& + &
\Res\limits_{y=0}\frac{1}{l_i-y}\left(\frac{\sqrt{y}}{\sqrt{l_i}}\jmath_<(y)+ \frac{y}{l_i}\theta_i \partial\psi_<(y)\right)
\left(\psi_>(y) + \frac{\sqrt{y}}{\sqrt{l_i}}\theta_i \jmath_>(y)\right).
\end{eqnarray}
Taking into account that
\begin{equation}
\psi_<(y)
=
\sum\limits_j\frac{\theta_j}{y-l_j},
\qquad
\jmath_<(y)  
=
\sum\limits_j\frac{\sqrt{l_j}}{\sqrt{y}}\frac{1}{l_j-y},
\end{equation}
we get
\begin{equation}
\frac{\sqrt{y}}{\sqrt{l_i}}\psi_<(y)+ \frac{y}{{l_i}}\theta_i \jmath_<(y)
=
\frac{\sqrt{y}}{l_i}\sum\limits_j\frac{\theta_j\sqrt{l_i}-\theta_i\sqrt{l_j}}{y-l_j},
\end{equation}
and thus
\begin{eqnarray}
\nonumber
&&
\hskip -2cm
\Res\limits_{y=0}\frac{1}{l_i-y}\left(\frac{\sqrt{y}}{\sqrt{l_i}}\psi_<(y)+ \frac{y}{l_i}\theta_i \jmath_<(y)\right)
\left(\jmath_>(y) + \frac{\sqrt{y}}{\sqrt{l_i}}\,\theta_i\partial\psi_>(y)\right)
\\[4pt]
& = & 
\frac{1}{l_i}\Res\limits_{y=0}\frac{\sqrt{y}}{l_i-y}\sum\limits_j\frac{\theta_j\sqrt{l_i}-\theta_i\sqrt{l_j}}{y-l_j}
\left(\jmath_>(y) + \frac{\sqrt{y}}{\sqrt{l_i}}\,\theta_i\partial\psi_>(y)\right)
\\[4pt]
\nonumber
& = &
\frac{1}{l_i}
\left(\Res_{y=l_i}+\Res_{y=l_j}\right)\sum\limits_j\frac{\sqrt{y}\left(\theta_j\sqrt{l_i}-\theta_i\sqrt{l_j}\right)}{(y-l_i)(y-l_j)}
\left(\jmath_>(y) + \frac{\sqrt{y}}{\sqrt{l_i}}\,\theta_i\partial\psi_>(y)\right)
\\[2pt]
\nonumber
& = & 
\sum\limits_{j;\,j\neq i}\frac{\theta_i\sqrt{l_j} - \theta_j\sqrt{l_i}}{l_i(l_i-l_j)}
\left(
\sqrt{l_i}\frac{\partial}{\partial l_i}
-
\sqrt{l_j}\frac{\partial}{\partial l_j}
\right),
\end{eqnarray}
where the simplicity of the resulting formula follows from the fact, that
\begin{eqnarray}
&&
\hskip -1cm
\left(\theta_j\sqrt{l_i}-\theta_i\sqrt{l_j}\right)
\left(\jmath_>(l_j)  + \frac{\sqrt{l_j}}{\sqrt{l_i}}\,\theta_i\partial\psi_>(l_j)\right)
\\[4pt]
\nonumber
& = & 
\left(\theta_j\sqrt{l_i}-\theta_i\sqrt{l_j}\right)\left(\jmath_>(l_j) +\theta_j\partial\psi_>(l_j)\right)
-
\left(\theta_j\sqrt{l_i}-\theta_i\sqrt{l_j}\right)\left(\theta_j
-\theta_i \frac{\sqrt{l_j}}{\sqrt{l_i}}\right)\partial\psi_>(l_j)
\\[2pt]
\nonumber
& = & 
-\left(\theta_j\sqrt{l_i}-\theta_i\sqrt{l_j}\right) \frac{\partial}{\partial l_j}+ 0.
\end{eqnarray}
Similar computation of the remaining term in  (\ref{eq:T_R_NS_new}) leads to the formula
\begin{eqnarray}
&&
\hskip -.5cm
\mathcal{T}^{\geqslant 0}_{\R\NS}(l_i,\theta_i) - \frac{\theta_i}{8l_i^2}
=
\frac{\partial}{\partial l_i}\left(\frac{\partial}{\partial \theta_i} + \theta_i\frac{\partial}{\partial l_i}\right)
+ \frac{\theta_i}{2l_i}\frac{\partial}{\partial l_i}
\\[2pt]
\nonumber
& + &
\sum\limits_{j;\,j\neq i}\frac{\theta_i\sqrt{l_j} - \theta_j\sqrt{l_i}}{l_i(l_i-l_j)}
\left(
\sqrt{l_i}\frac{\partial}{\partial l_i}
-
2\sqrt{l_j}\frac{\partial}{\partial l_j}
\right)
+
\sum\limits_{j;\,j\neq i}\left(\frac{\sqrt{l_j}}{\sqrt{l_i}(l_i-l_j)} + \frac{\theta_i\theta_j}{(l_i-l_j)^2}\right)
\left(D_i-D_j\right).
\\[4pt]
\nonumber
& = &
\mT^{\geqslant -\frac12}_{\NS\NS}(l_i,\theta_i) +
\delta \mathcal{T}^{\geqslant 0}_{\R\NS}(l_i,\theta_i)
\end{eqnarray}
with
\begin{equation}
\delta\mathcal{T}^{\geqslant 0}_{\R\NS}(l_i,\theta_i)
=
\frac{2\theta_i}{l_i}\sum\limits_j\frac{\partial}{\partial l_j}
-
\frac{1}{\sqrt{l_i}}\sum\limits_j\frac{1}{\sqrt{l_i}+\sqrt{l_j}}
\left(\frac{\partial}{\partial \theta_i} 
- \frac{\partial}{\partial \theta_j}
+ 2\theta_i\frac{\partial}{\partial l_i}
+  \theta_j\frac{\partial}{\partial l_j}\right)
\end{equation}
Notice that, in contrast to the R-R case, $\delta\mathcal{T}^{\geqslant 0}_{\R\NS}(l_i,\theta_i)$
contains derivatives with respect to $l_j$ and $\theta_j.$
\subsubsection{The NS-R sector}
We have
\begin{equation}
\mJ(\vt,x) = \vt\jmath_{\NS}(x) + \psi_\R(x) = \sum \limits_{m\in\mathbb Z}\left(\vt\sa_m x^{-m-1} + \psi_m x^{-m-\frac12}\right)
\end{equation}
and, in order to represent the mode algebra
\begin{equation}
\left[\sa_m,\sa_n\right] = m\delta_{m+n,0},
\qquad
\left\{\psi_m,\psi_n\right\} = \delta_{m+n,0},
\end{equation}
we take
\begin{equation}
\sa_m = \frac{\partial}{\partial g_m}, \quad \psi_m = \frac{\partial}{\partial q_m},
\quad
\sa_{-m} = m g_m, \quad \psi_{-m} = q_m, \quad m >  0,
\qquad
\sa_0 = \frac{\partial}{\partial g_0},
\quad
\psi_0 = \frac{q_0}{2} + \frac{\partial}{\partial q_0}.
\end{equation}
We next decompose
\begin{equation}
\jmath_\NS(x) = \jmath_{\geqslant}(x)+ \jmath_{<}(x),
\qquad
\jmath_{\geqslant}(x) = \sum\limits_{m\geqslant 0}x^{-m-1}\frac{\partial}{\partial g_m},
\quad
\jmath_<(x) = \sum\limits_{m\geqslant 1}mg_m x^{m-1},
\end{equation}
as well as
\begin{equation}
\psi_\R(x) = \psi_\geqslant(x)+ \psi_{<}(x),
\qquad
\psi_\geqslant(x) = \sum\limits_{m\geqslant 0}x^{-m-\frac12}\frac{\partial}{\partial q_m},
\qquad
\psi_<(x) = \frac{q_0}{2\sqrt{x}} + \sum\limits_{m\geqslant 1}q_m x^{m-\frac12}.
\end{equation}
Since
\begin{equation}
\left\{\psi_\geqslant(y),\psi_<(x)\right\} = \frac{y+x}{2\sqrt{y}\sqrt{x}(y-x)} = \frac{1}{y-x} + \frac{y-x}{8x^2} + \mathcal{O}\left((y-x)^2\right),
\quad
\left[\jmath_\geqslant(y),\jmath_<(x)\right] = \frac{1}{(y-x)^2},
\end{equation}
and there is neither singularity of the order $x^{-2}$ in $\jmath_<(x)\jmath_<(x) + \partial\psi_<(x)\psi_<(x)$ nor singularity of order $x^{-\frac32}$ in 
$\jmath_<(x)\psi_<(x),$ then proceeding as in the R-NS sector we get

\begin{eqnarray}
\label{eq:T_R_NS_1}
\nonumber
\mT^{\geqslant 0}_{\NS\R}(l_i,\theta_i) - \frac{\theta_i}{8 l_i^2}
& = &
\left(\jmath_\geqslant(x) + \theta_i\partial\psi_\geqslant(x)\right)\left(\theta_i\jmath_\geqslant(x) + \psi_\geqslant(x)\right)\Big|_{x=l_i}
\\[4pt]
& + &
\Res\limits_{y=0}\frac{1}{l_i-y}\left(\frac{\sqrt{y}}{\sqrt{l_i}}\psi_<(y)+ \frac{y}{l_i}\theta_i \jmath_<(y)\right)
\left(\jmath_\geqslant(y) + \frac{\sqrt{y}}{\sqrt{l_i}}\,\theta_i\partial\psi_\geqslant(y)\right)
\\[4pt]
\nonumber
& + &
\Res\limits_{y=0}\frac{1}{l_i-y}\left(\frac{\sqrt{y}}{\sqrt{l_i}}\jmath_<(y)+ \frac{y}{l_i}\theta_i \partial\psi_<(y)\right)
\left(\psi_\geqslant(y) + \frac{\sqrt{y}}{\sqrt{l_i}}\theta_i \jmath_\geqslant(y)\right).
\end{eqnarray}
The super-Miwa transformation now takes the form:
\begin{equation}
g_0 = -\sum\limits_j\log l_j, \quad g_m = \frac{1}{m}\sum\limits_j l_j^{-m}, \quad m > 0,
\qquad
q_n = -\sum\limits_{j}\theta_j l_j^{-n-\frac12}, \quad n \geqslant 0,
\end{equation}
so that we preserve the crucial (in our approach defining) relations
\begin{equation}
\psi_\geqslant(x)\Big|_{x=l_i}
=
-\frac{\partial}{\partial\theta_i},
\qquad
\left(\jmath_\geqslant(x) + \theta_i\partial\psi_\geqslant(x)\right)\Big|_{x=l_i}
=
-\frac{\partial}{\partial l_i}
\end{equation}
and 
\begin{equation}
\label{eq:NS_R_first_contribution}
\left(\jmath_\geqslant(x) + \theta_i\partial\psi_\geqslant(x)\right)\left(\theta_i\jmath_\geqslant(x) + \psi_\geqslant(x)\right)\Big|_{x=l_i}
=
\frac{\partial}{\partial l_i}\left(\frac{\partial}{\partial \theta_i} + \theta_i\frac{\partial}{\partial l_i}\right) + \partial\big(\theta_i\jmath_\geqslant(x) + \psi_\geqslant(x)\big)\Big|_{x=l_i}.
\end{equation}
Using equalities
\begin{equation}
j_<(y) = \sum\limits_j\frac{1}{l_j-y}, 
\qquad
\psi_<(y) = -\sum\limits_j\frac{\sqrt{y}}{\sqrt{l_j}}\frac{\theta_j}{l_j-y} + \frac{q_0}{2\sqrt{y}},
\end{equation}
we obtain an expression for $\mathcal{T}^{\geqslant 0}_{\NS\R}(l_i,\theta_i)$ of the form:
\begin{align}
\mT_{\NS\R}^{\geqslant0}(l_i,\theta_i)
-\frac{\theta_i}{8l_i^2}
={}&
D_i^3
+
\frac{1}{2l_i}
\left(
D_i+\theta_iD_i^2
\right)
\nonumber\\
&+
\sum_{j;\,j\neq i}
\frac{1}{l_i-l_j}
\Bigg[
D_i
-
\sqrt{\frac{l_j}{l_i}}\,D_j
+
\left(
\theta_i-\sqrt{\frac{l_i}{l_j}}\,\theta_j
\right)
\left(
\frac{\partial}{\partial l_i}
-
2\frac{l_j}{l_i}\frac{\partial}{\partial l_j}
\right)
\Bigg]
\nonumber\\
&+
\sum_{j;\,j\neq i}
\frac{\theta_i\theta_j}{(l_i-l_j)^2}
\left[
\frac{l_i+l_j}{2\sqrt{l_i l_j}}\,
\frac{\partial}{\partial\theta_i}
-
\frac{\partial}{\partial\theta_j}
\right]
-
\frac{q_0}{2\sqrt{l_i}}
\left(
\frac{\partial}{\partial l_i}
+
\frac{\theta_i}{2l_i}
\frac{\partial}{\partial\theta_i}
\right).
\label{eq:NS_R_final_operator}
\end{align}

\section{Airy structures}
\label{sec:airy}
Differential operators constructed in the previous section contain information about the behavior of free,
quantum fields on a Riemann surface around a branch-point singularity of a local coordinate map. We can think of $x$ as parameterizing a disc centered in a point ``infinitely distant'' from the rest of the surface. By this we mean that formulas like (\ref{eq:bosonc_current_on_x_plane}) or (\ref{eq:super_RR_current_on_x_plane}) are valid for
any finite $x$ and the information on the global geometry of the Riemann surface is encoded in the ``bra'' $\bra{V}$ state at infinity \cite{Kostov10,KO10}. More precisely, the geometry of the Riemann surface
can be extracted form the classical part of the $j$ current, that is, from its vacuum expectation value \cite{KO10}. Notice now
that for any $\mu_t\in \mathbb{C}$ with $t\in \mathbb{Z}_{>0}$ or $t\in \mathbb{Z}_{>0}+\frac12,$ if
\begin{equation}
[\sa_t,\sa_s] = t\delta_{t+s,0}
\end{equation}
then
\begin{equation}
\bra{0}\sum\limits_{t > 0}
 \mu_t x^{t-1} = \bra{0}\sum\limits_{t > 0} (\mu_t + \sa_{-t})x^{t-1}
=
\bra{0}\text{e}^{\sum\limits_{t>0}\frac{\mu_t}{t}\sa_t}\sum\limits_{s > 0}\sa_{-s}x^{s-1}\text{e}^{-\sum\limits_{t >0}\frac{\mu_t}{t}\sa_t}
\end{equation}
and we see that we can control the behavior of the classical current at infinity by choosing $\bra{V}$ to be 
a coherent state
\begin{eqnarray}
\bra{V} = \bra{0}\exp\Big\{\sum\limits_{t>0}\mu_t\frac{\sa_t}{t}\Big\}.
\end{eqnarray}
CFT approach we are advocating in the present work is  in a natural way related to the methods of topological recursion \cite{KO10}. Namely, let
$\mathcal S=(\Sigma,x,\omega_{0,1},\omega_{0,2})$ be a spectral curve (c.f.\ for instance \cite{EO07,Bouchard24})) where, in the local $z$ coordinate the one-form  $\omega_{0,1}$ can be expressed as
$\omega_{0,1}=y(z)\,dx(z).$  The macroscopic potential $V(z)$ and operator $J[V]$
are then defined as
\begin{equation}
    V(z) = \Res_{z'\to 0}\,\log\!\Big(1-\frac{z}{z'}\Big)\,y(z')\,dx(z'), \hspace{1cm} J[V] = -\oint_{\infty}\frac{dz}{2\pi i}\,V(z)\,j(z).
\end{equation}
In this paragraph we use the local normalisation
$x(z)=z^2/2$, so that at the rank-two ramification point $z=0$ one has
$dx(z)=z\,dz$ and thus $\omega_{0,1} = y(z) z\,dz.$
Expanding $y(z)=\sum_m \mu_{m+2} z^m$ we get
\begin{equation}
    V(z)
    = \Res_{z'\to0}\log\!\Big(1-\frac{z}{z'}\Big)z'y(z')\,dz'
    = -\sum_{n=1}^{\infty}\frac{1}{n}z^n\Res_{z'\to0}(z')^{1-n}y(z')\,dz'
    = -\sum_{n=-1}^{\infty}\frac{\mu_{n+2}}{n+2}\,z^{n+2},
\end{equation}
so that
\begin{equation}
    J[V]
    = \sum_{n=-1}^{\infty}\frac{\mu_{n+2}}{n+2}\oint_\infty\frac{dz}{2\pi i}\,z^{n+2}j(z)
    = \sum_{n=-1}^{\infty}\mu_{n+2}\frac{a_{n+2}}{n+2}
    = \sum_{n=1}^{\infty}\mu _{n}\frac{a_n}{n}.
\end{equation}
We see that  a coherent-state deformation of the Fock we need is generated  through
the right action of the operator $\exp J[V]$ on the ``bra'' vacuum $\bra{0},$ 
\begin{equation}
\bra{V}
=
\bra{0}\exp J[V]
\end{equation}
while
\begin{equation}
\omega_{0,1}(z)
=
y(z)dx(z)
=
\bra{V}j(z)\ket{0}dz.
\end{equation}

To avoid a possible confusion let us notice that,
in the terminology of \cite{Bouchard24}, a linear potential $V$ corresponds to the Bessel curve and cubic
potential to the Airy curve.
In the remainder of this section, however, the term
``Airy structure'' is used in the algebraic Kontsevich--Soibelman sense.
The shifts below act on the $x$-plane modes $\sa_m$, whereas the preceding
Bessel/Airy dictionary is formulated for the potential $V(z)$ and the
$z$-plane modes $a_n$.  Since the two descriptions are related by a
ramified coordinate $x\propto z^2$ and by sector-dependent mode normalisations,
the label of a shifted mode $\sa_m$ should not by itself be interpreted as
the degree of the corresponding potential $V(z)$.

In order to describe a general choice of polarization, we introduce the quadratic
operator
\begin{equation}
\Phi_{\mathrm{bos}}
=
\exp\left(
\sum_{m,n\geq1}
\frac{\varphi_{mn}}{2mn}\,\sa_m\sa_n
\right)
\end{equation}
and the corresponding polarised bra-state
\begin{equation}
\big\langle\hskip1.5pt\widetilde V\,\big|
=
\bra{0}\exp J[V]\,\Phi_{\mathrm{bos}}.
\end{equation}
Since $\Phi_{\mathrm{bos}}$ also contains only positive modes,
$\big\langle\hskip1.5pt\widetilde V\,\big| \big|\,0\,\big\rangle=1$.  Moreover, the quadratic deformation does
not contribute to the one-point function, so that
\begin{equation}
\big\langle\hskip1.5pt\widetilde V\,\big|j(z)\big|\,0\,\big\rangle
=
\bra{V}j(z)\ket{0}.
\end{equation}
The second differential entering the initial data of topological recursion
is the fundamental bidifferential $\omega_{0,2}$, usually referred to as the
Bergman kernel.  From the CFT perspective, it is the connected two-point
function $\langle j(z)j(w)\rangle_{c}$ of the bosonic current:
\begin{align}
\omega_{0,2}(z,w)
={}&
\left[
\frac{\big\langle\hskip1.5pt\widetilde V\,\big|j(z)j(w)\big|\,0\,\big\rangle}
     {\big\langle\hskip1.5pt\widetilde V\,\big|\,0\,\big\rangle}
-
\frac{
\big\langle\hskip1.5pt\widetilde V\,\big|\,j(z)\big|\,0\,\big\rangle\,
\big\langle\hskip1.5pt\widetilde V\,\big|\,j(w)\big|\,0\,\big\rangle
}{
\big\langle\hskip1.5pt\widetilde V\,\big|\,0\,\big\rangle^{\,2}
}
\right]dz\,dw
\nonumber\\
={}&
\left(
\frac{1}{(z-w)^2}
+
\sum_{m,n\geq1}
\varphi_{mn}z^{m-1}w^{n-1}
\right)dz\,dw.
\end{align}
The double pole is fixed by the local current OPE, whereas the regular
part is determined by the choice of polarisation.  Its symmetric
coefficients $\varphi_{mn}$ are the polarisation coefficients.  Thus,
the linear operator $J[V]$ determines $\omega_{0,1}$, while the quadratic
operator $\Phi_{\mathrm{bos}}$ determines the regular part of
$\omega_{0,2}$.

Let us at this point recall a definition of the Airy structure, introduced by Maxim Kontsevich and Ian Soibelman
in \cite{KS17} and further extended to the graded (or super) case in \cite{BCHORS20} (see also \cite{Bouchard24} for a pedagogical introduction).

A (super) quantum Airy structure is a collection of differential operators (Hamiltionians) $\{H_s\}_{s \in I}$ in $\mathbb{Z}_2$-graded variables $\{x_s\}_{s \in I}$, which take the form:
\begin{equation}
    H_s = \hbar \frac{\partial}{\partial x_s} + Q_s + \hbar d_s,
\end{equation}
where, in the simplest case, $Q_s$ are purely quadratic polynomials in the variables $x_t$ and derivatives $\hbar \frac{\partial}{\partial x_t}$, and $d_s \in \mathbb{C}.$ The $\mathbb{Z}_2$-grading of the operator $H_s$ equals the grading of the corresponding variable $x_s$. Furthermore, the collection satisfies a graded Lie algebra condition:
$  [H_s, H_t]_{\pm} = \hbar \sum_{u \in I} f_{st}^u H_u,$
where $f_{st}^u \in \mathbb{C}$ are structure constants, and $[ \cdot, \cdot ]_{\pm} = H_s H_t - (-1)^{|H_s||H_t|} H_t H_s$ denotes the graded commutator.

The presence in each $H_s$ of a linear term is crucial in a proof of the existence and uniqueness of the partition function $Z,$  defined by equations $H_s Z=0,\; s \in I.$

Operators $L_m$ in the bosonic case, or $L_m$  and $G_s$ in the graded case form a (graded) Lie algebra, but they do not form an Airy structure.
These operators are either purely quadratic in the free field modes (in the NS case) or 
are completed with the constant term (in the Ramond case), and in the representation discussed in Section \ref{sec:section2} the free field modes are indeed multiplication/differentiation operators, but
$L_m$ and $G_s$ discussed in this section lack the linear term. Notice, however, that for $m\geqslant -1$
\begin{equation}
\exp\Big\{\sum\limits_{n\geqslant 1}\mu_n\frac{\sa_n}{n}\Big\}
L_{\NS,m}
\exp\Big\{-\sum\limits_{n\geqslant 1}\mu_n\frac{\sa_n}{n}\Big\}
=
L_{\NS,m} + \sum\limits_{n\geqslant 1} \mu_n \sa_{m+n}
\; \to \; 
L_{\NS,m} + \sum\limits_{n\geqslant 1} \mu_{n} \frac{\partial}{\partial g_{m+n}}
\end{equation}
and this similarity transformation does not change the Virasoro algebra.

In the simplest case of $\mu_n = \delta_{n,1}$ we thus directly obtain the Airy structure (with the parameter $\hbar$ set to 1). In the general case we  have to diagonalize operators $\sum\limits_{n\geqslant 1} \mu_{n} \frac{\partial}{\partial g_{m+n}}.$ Since the set of equations
\begin{equation}
\frac{\partial}{\partial p_{m+1} } = \sum\limits_{n\geqslant 0} \mu_{n+1} \frac{\partial}{\partial g_{m+n+1}}
\end{equation}
is triangular, i.e.\ $\partial/\partial p_n$ depends only on $\partial/\partial g_m$ with $m\geqslant n,$ it can be explicitly inverted for any finite set of non-zero $\mu_n$-s. For instance, if the only non-zero elements are $\mu_1$ and $\mu_2$ then we get
\begin{equation}
g_0 = \mu_1 p_0, \quad g_m = \mu_1 p_m + \mu_2 p_{m-1},\;\; m \geqslant 1,
\qquad
\mu_1\frac{\partial}{\partial g_m} = \sum\limits_{n\geqslant 0} \left(-\frac{\mu_2}{\mu_1}\right)^n\frac{\partial}{\partial p_{m+n}}, \;\; m \geqslant 0.
\end{equation}
This is a linear transformation and the resulting set of differential operators in the variables $p_m$ forms an Airy structure.

Let us now discuss Kontsevich operators corresponding to the (simplest) Airy structures for the cases presented in the previous sections.
\begin{itemize}
\item In the bosonic, NS case discussed above, with $\mu_n = \mu \delta_{n,1}$ we get
\begin{equation}
\sT^{\geqslant -1}_{\NS,\Ai}(x)
:=
\text{e}^{\mu\sa_{1}}\sT^{\geqslant -1}_{\NS}(x)\text{e}^{-\mu\sa_{1}}
=
\mu \sum\limits_{m\geqslant 0}\frac{\sa_m}{x^{m+1}} + \sT^{\geqslant -1}_{\NS}(x) 
=
\mu\sj_{\NS,\geqslant}(x) + \sT^{\geqslant -1}_{\NS}(x),
\end{equation}
and consequently
\begin{equation}
\sT^{\geqslant -1}_{\NS,\Ai}(l_i) = - \mu \frac{\partial}{\partial l_i} + \sT^{\geqslant -1}_{\NS}(l_i) 
=
- \mu \frac{\partial}{\partial l_i}
+
\frac{\partial^2}{\partial l_i^2} + \sum\limits_{j;\,j\neq i} \frac{1}{l_i-l_j}\left(\frac{\partial}{\partial l_i}- \frac{\partial}{\partial l_j}\right).
\end{equation}
\item 
In the bosonic, Ramond case we take $\mu_k = \mu \delta_{k,\frac32}$ so that
\begin{equation}
\sT^{\geqslant -1}_{\R,\Ai}(x)
:=
\text{e}^{\frac23\mu\sa_{3/2}}\sT^{\geqslant -1}_{\R}(x)\text{e}^{-\frac23\mu\sa_{3/2}}
=
\mu \sum\limits_{k\geqslant \frac12}\frac{\sa_k}{x^{k+\frac12}} + \sT^{\geqslant -1}_{\R}(x) 
=
\mu\sqrt{x}\sj_{\R,>}(x) + \sT^{\geqslant -1}_{\R}(x),
\end{equation}
and in effect
\begin{eqnarray}
\label{eq:Airy_structure_R_bosonic}
\sT^{\geqslant -1}_{\R,\Ai}(l_i) 
& = &
- \mu \sqrt{l_i} \frac{\partial}{\partial l_i} + \sT^{\geqslant -1}_{\R}(l_i) 
=
\frac{\partial^2}{\partial l_i^2} +\!\! \sum\limits_{j;\,j\neq i} \frac{1}{l_i-l_j}\left(\frac{\partial}{\partial l_i}- \frac{\partial}{\partial l_j}\right)
\\[4pt]
\nonumber
&&
-
\left(\mu\sqrt{l_i} +\frac{1}{\sqrt{l_i}}\sum\limits_j\frac{1}{\sqrt{l_i}+\sqrt{l_j}}\right)\frac{\partial}{\partial l_i}
+
\frac{1}{4l_i}\Big(\sum\limits_j\frac{1}{\sqrt{l_j}}\Big)^2 \!\!
+
\frac{1}{16 l_i^2}.
\end{eqnarray}
\item 
In the supersymmetric, NS-NS case, we need
to shift $L_m$ with $m\geqslant {-1}$ by $\partial/\partial g_{m+1}.$ This is
achieved using, as above, conjugation with the $\sa_1$ mode of the scalar field. In effect
\begin{eqnarray}
\mT^{\geqslant -\frac12}_{\NS\NS,\Ai}(x,\vt)
:=
\text{e}^{\mu\sa_{1}}\mT^{\geqslant -\frac12}_{\NS\NS}(x,\vt)\text{e}^{-\mu\sa_{1}}
=
\mu\left(\mJ_>(x,\vt) +\vt \mDJ_\geqslant(x,\vt)\right)+\mT^{\geqslant -\frac12}_{\NS\NS}(x,\vt) 
\end{eqnarray}
so that
\begin{equation}
\mT^{\geqslant -\frac12}_{\NS\NS,\Ai}(l_i,\theta_i)
=
-\mu \big(D_i +\theta_iD_i^2\big) +D_i^3
+
\sum\limits_{j;\,j\neq i}\left(
\frac{\theta_i-\theta_j}{l_i-l_j} \left(\frac{\partial}{\partial l_i} - 2\frac{\partial}{\partial l_j}\right)
+
\frac{D_i - D_j}{l_i-l_j-\theta_i\theta_j}\right).
\end{equation}
\item 
In the R--R case, in order for the linear term to be composed of
derivatives with respect to all bosonic variables
$g_k,\; k\geqslant\frac12$, we need to shift the generators
$L_n,\; n\geqslant-1$, by a term proportional to
$\partial/\partial g_{n+\frac32}$.  Denoting the corresponding dilaton
parameter by $\mu$, we obtain
\begin{eqnarray}
\label{eq:Airy_structure_RR_temp}
\nonumber
\mT^{\geqslant-\frac12}_{\R\R,\Ai}(x,\vt)
&:=&
\text{e}^{\frac23\mu\sa_{3/2}}
\mT^{\geqslant-\frac12}_{\R\R}(x,\vt)
\text{e}^{-\frac23\mu\sa_{3/2}} =
\mu\sqrt{x}
\left(
\sum\limits_{m\geqslant1}
\frac{\psi_m}{x^{m+\frac12}}
+
\sum\limits_{k\geqslant\frac12}
\frac{2\vt\sa_k}{x^{k+1}}
\right)
+
\mT^{\geqslant-\frac12}_{\R\R}(x,\vt)
\\[4pt]
&=& \mu
\sqrt{x}
\big(
\mJ_>(x,\vt)+\vt\mDJ_>(x,\vt)
\big)
-
\mu\frac{\partial}{\partial q_0}
+
\mT^{\geqslant-\frac12}_{\R\R}(x,\vt),
\end{eqnarray}
so that
\begin{equation}
\label{eq:Airy_structure_RR}
\mT^{\geqslant-\frac12}_{\R\R,\Ai}(l_i,\theta_i)
=
-\mu\sqrt{l_i}
\big(
D_i+\theta_iD_i^2
\big)
-
\mu\frac{\partial}{\partial q_0}
+
\mT^{\geqslant-\frac12}_{\R\R}(l_i,\theta_i).
\end{equation}
As is evident from the first line of (\ref{eq:Airy_structure_RR_temp}) (recall that $\psi_m = \partial/\partial q_m,\; m \geqslant 1$) the linear, proportional to $\mu$ term
in (\ref{eq:Airy_structure_RR}) does not contain the derivative with respect to $q_0.$ The Ramond zero mode is nevertheless present since the operator $\mT^{\geqslant-\frac12}_{\R\R}(l_i,\theta_i)$  
depends on it. This is a Super Quantum Airy Structure with one
additional odd coordinate in the precise sense of Definition~2.3 of
\cite{BCHORS20}.  More explicitly, the Hamiltonians are indexed by the
variables $g_k$, $k\geqslant\frac12$, and $q_m$, $m\geqslant1$, but act on the Weyl algebra of
the enlarged space
$\widetilde V=V\oplus\mathbb K^{0|1}$, which also contains multiplication
by $q_0$ and differentiation by $\partial_{q_0}$.  There is no separate
Hamiltonian indexed by $q_0$.

The existence-and-uniqueness statement of Theorem~2.10 of
\cite{BCHORS20} applies to this enlarged setting.  Indeed, writing the free energy $\mathcal{F} = \log Z$ as 
$\mathcal{F}=\mathcal{F}^{(0)}+q_0\mathcal{F}^{(1)}$, one has
$\partial_{q_0}\mathcal{F}=\mathcal{F}^{(1)}$, so the appearances of
$\partial_{q_0}\mathcal{F}$ in the quadratic part of the constraints are
determined recursively together with the mixed coefficients containing
$q_0$.  After all indexed constraints have been imposed, a possible
residual ambiguity can depend only on $q_0$ and therefore has the form
$a+bq_0$.  We impose the standard normalisation
$\mathcal F(0)=0$ and require the free energy to be even; these are
normalisation conditions on the partition function, rather than
consequences of the constraints, and they exclude both terms.  This does
not make the free energy independent of $q_0$: mixed even monomials such
as $q_0q_m$ are allowed and are fixed by the Airy recursion.

\item 
Let us recall that in both R-NS and NS-R sectors we are dealing with a subset of the Ramond super-algebra formed by operators
$L_n - \frac{1}{16}\delta_{n,0}, G_n,\; n \geqslant 0.$ In the R-NS sector we realize these operators through multiplication and differentiation with respect to a set of bosonic variables $g_k,$ and fermionic $q_k,$  $k \in \mathbb{Z}_{\geqslant 0} + \tfrac12.$ To form an Airy structure we thus need to shift $L_n$ by linear term containing a derivative with respect to $g_{n+\frac12}$ and
$G_n$ by a term proportional to $q_{n+\frac12}.$ This is achieved through the similarity transformation 
\begin{equation}
\mT^{\geqslant 0}_{\R\NS,\Ai}(x,\vt)
:= 
\text{e}^{2\mu\sa_{1/2}}\mT^{\geqslant 0}_{\R\NS}(x,\vt)\text{e}^{-2\mu\sa_{1/2}}
=
\frac{\mu}{\sqrt{x}}\big(\psi_>(x) + 2\vt\jmath_>(x)\big) +  \mT^{\geqslant 0}_{\R\NS}(x,\vt)
\end{equation}
and thus
\begin{equation}
\mT^{\geqslant 0}_{\R\NS,\Ai}(l_i,\theta_i)
=
-\frac{\mu}{\sqrt{l_i}} \big(D_i +\theta_i D_i^2\big)+ \mT^{\geqslant 0}_{\R\NS}(l_i,\theta_i).
\end{equation}
Since there is no zero mode in the fermionic sector, we end up with an example of a ``regular'' Super Quantum Airy Structure.
\item  
Finally, in the NS-R sector we are dealing with a set of bosonic variables $g_m$ and fermionic variables $q_m,\; m \in \mathbb{Z}_{\geqslant 0}.$ We can achieve a shift of an operator $L_n$ by a term proportional to $\partial/\partial g_n$ by shifting the mode $\sa_0 = \partial/\partial g_0.$ This can be realized through a conjugation with an operator $\exp\big\{-\mu g_0\big\}.$ In effect
\begin{eqnarray}
\nonumber
\mT^{\geqslant 0}_{\NS\R,\Ai}(x,\vt)
& := &
\text{e}^{-\mu g_0}\mT^{\geqslant 0}_{\NS\R}(x,\vt)\text{e}^{\mu g_0}
=
\frac{\mu^2 \vt}{x^2}
+
\frac{\mu}{x}\sum\limits_{m\geqslant 0}\left(\frac{\psi_m}{x^{m+\frac12}} + \frac{2\vt a_m}{x^{m+1}}\right)
+
\mT^{\geqslant 0}_{\NS\R}(x,\vt)
\\[2pt]
& = &
\frac{\mu^2 \vt}{x^2}
+
\frac{\mu}{x}\big(\psi_\geqslant(x) + 2\vt\jmath_\geqslant(x)\big) + \frac{\mu}{2\sqrt{x^3}}q_0 +
\mT^{\geqslant 0}_{\NS\R}(x,\vt),
\end{eqnarray}
and
\begin{equation}
\mT^{\geqslant 0}_{\NS\R,\Ai}(l_i,\theta_i) - \frac{\theta_i}{8l_i^2}
=
\left(\mu^2-\tfrac18\right)\frac{\theta_i}{l_i^2}
-\frac{\mu}{l_i}\big(D_i +\theta_i D_i^2\big) + \frac{\mu}{2l_i^{3/2}}q_0 + \mT^{\geqslant 0}_{\NS\R}(l_i,\theta_i).
\end{equation}
Due to the presence of a linear term containing the fermionic zero mode $q_0$ this is not an Airy structure. Its presence is a result of the relation
\begin{equation}
    \delta G_0 = \text{e}^{-\mu g_0}G_0\text{e}^{\mu g_0} -G_0 = \mu\psi_0 = \mu\partial_{q_0}+\frac{\mu}{2}q_0,
\end{equation}
The fermionic zero mode may be removed by excluding $G_0$
from the constructed Airy structure.  In that case the relation
$G_0^2=L_0-\tfrac{1}{16}$ is no longer imposed, and the constant in the
lowest bosonic constraint is not fixed.  One may therefore consistently
supplement the retained generators with
$(L_0-C)Z=0$ for an arbitrary constant $C$.  This gives a one-parameter
family of Airy structures, each with a uniquely determined partition
function, rather than a non-unique partition function for a fixed Airy
structure.

As an alternative, we can define the NS-R Airy structure by shifting (as in the NS-NS case) the $\sa_1$ mode. As a result
\begin{equation}
\delta L_n = \mu\frac{\partial}{\partial g_{n+1}}, \quad \delta G_n=\mu\frac{\partial}{\partial q_{n+1}}, \quad n \geqslant 0,
\end{equation}
and
\begin{equation}
\mT^{\geqslant 0}_{\NS\R,\Ai}(x,\vt)
:=
\text{e}^{\mu\sa_{1}}\mT^{\geqslant 0}_{\NS\R}(x,\vt)\text{e}^{-\mu\sa_{1}}
=
\mu\left(\psi_\geqslant(x) - \frac{\partial_{q_0}}{\sqrt{x}}\right)
+
2\mu\vt\left(\jmath_\geqslant(x) - \frac{\partial_{g_0}}{x}\right)
+
\mT^{\geqslant 0}_{\NS\R}(x,\vt),
\end{equation}
so that
\begin{equation}
\mT^{\geqslant 0}_{\NS\R,\Ai}(l_i,\theta_i)
=
-\mu\big(D_i +\theta_i D_i^2\big)
-\frac{\mu}{\sqrt{l_i}}\frac{\partial}{\partial q_0}
-\frac{2\mu\theta_i}{l_i}\frac{\partial}{\partial g_0}
+ \mT^{\geqslant 0}_{\NS\R}(l_i,\theta_i).
\end{equation}
These operators do not determine the  $g_0$ dependence of the partition function, but since $\sa_0$ is a central element we can  consistently supplement the set of equations satisfied by $Z$ with an additional
constraint $\sa_0 Z = \frac{\partial}{\partial g_0}Z = 0$ and deal with the  fermion zero mode problem as in the RR case.
\end{itemize}

More generally, the dilaton shift may be combined with a change of
polarisation.  In the NS--NS sector these transformations are achieved by inserting a state
\begin{eqnarray}
\big\langle\,\tilde{V}\,\big| = \bra{0}\exp\Big\{\sum\limits_{t>0}\mu_t\frac{\sa_t}{t} + 
\sum_{m,n\geq1}
\frac{\varphi_{mn}}{2mn}\,\sa_m \sa_n
+
\sum_{r,s\geq\frac12}
\frac{\chi_{r,s}}{2}\,\psi_r\psi_s
 \Big\}.
 \label{eq:NS_NS_shift_polarization}
\end{eqnarray}
Here $\mu_k$ specify the dilaton background, while
$\varphi_{mn}$ and $\chi_{r,s}$ are the bosonic and fermionic
polarisation coefficients, respectively.  In particular, the latter
appear in the regular part of the fermionic two-point function (Szeg\H{o} Kernel), 
\begin{equation}
\big\langle\,\tilde{V}\,\big|
\psi(z)\psi(w)
\big|\hskip 1pt0\hskip 1pt \big\rangle
=
\frac{1}{z-w}
+
\sum_{r,s\geq\frac12}
\chi_{s,r}\,
z^{r-\frac12}w^{s-\frac12},
\label{eq:fermionic-polarization-correlator}
\end{equation}
with $\chi_{r,s}=-\chi_{s,r}$. Analogous transformations can be defined in the remaining sectors by
replacing the mode sets in \eqref{eq:NS_NS_shift_polarization} according to
the corresponding bosonic and fermionic monodromies.  For more general backgrounds, especially when negative
modes are included, the conjugation became more complicated, as in \cite{AKOO25}. We do not consider this
extension here.

We conclude this section by placing the free super-current construction
in the context of two related approaches to supersymmetric Virasoro
constraints.  The use of a single super-current $\mJ(x,\theta)$ is natural,
since its quadratic tensor
\begin{equation}
\mT(x,\theta)
=
:\mathcal D\mJ(x,\theta)\mJ(x,\theta):
=
:j(x)\psi(x):
+
\theta
\left(
:j(x)^2:
+
:\partial\psi(x)\psi(x):
\right)
\label{eq:supercurrent-quadratic-tensor}
\end{equation}
packages the fermionic superconformal current $G(z)$ and the energy-momentum
tensor $T(z)$ into a single superfield.  It treats uniformly the four independent
monodromy assignments considered here and is compatible with the
sector-dependent super-Miwa representations derived in
Section~\ref{sec: Section 3}.

The first related framework is the $\mathcal N=1$ super-topological
recursion of \cite{BO21}.  It reformulates the abstract super loop equations
in two equivalent ways: as a recursion on a local super spectral curve and
as a system of differential constraints forming a Super Quantum Airy
Structure.  Our construction is complementary.  Starting directly from
free bosonic and fermionic currents, we keep their monodromies independent
and obtain the four NS--NS, NS--R, R--NS, and R--R operator realisations.
In the local NS--NS setting, and for trivial additional polarisation, the
resulting constraints belong, after matching the mode and covering
conventions, to the same local $\mathcal N=1$ super-Virasoro/Airy framework
as those of \cite{BO21}.  The super-Miwa transformations provide explicit differential representations of these constraints in the external spectral variables.

A more direct geometric comparison is provided by the superconformal
topological recursion of \cite{AKOO25}, formulated on a super-Riemann
surface with a Ramond divisor.  For an even differential
$\omega=y[dx]+\lambda[d\varphi]$, its super quadratic Casimir has the local
form
\begin{equation}
\mathcal C\!\left(\omega^{\boxtimes 2}\right)
=
[d\varphi]^3
\left(
C_1+\varphi C_0
\right),
\qquad
C_1=y\lambda,
\qquad
C_0=y^2+\lambda'\lambda.
\label{eq:super-quadratic-Casimir}
\end{equation}
The local fields are related to our CFT variables by
\begin{equation}
y\longleftrightarrow j,
\qquad
\lambda\longleftrightarrow\psi,
\qquad
\varphi\longleftrightarrow\theta.
\label{eq:Casimir-current-dictionary}
\end{equation}
After quantisation and normal ordering, the two components are identified as
\begin{equation}
C_1+\theta C_0
\longleftrightarrow
:j\psi:
+
\theta
\left(
:j^2:
+
:\partial_x\psi\,\psi:
\right)
=
\mT(x,\theta).
\label{eq:Casimir-supercurrent-identification}
\end{equation}
Thus, up to the differential-form factor and normalisation conventions, the
super quadratic Casimir contains the same two local components as our
quadratic super-current tensor.  The projected Hamiltonians of
\cite{AKOO25} additionally involve the twisted Casimir, the local
involution, and the fundamental bidifferential.  The comparison above is
therefore restricted to the local quadratic density.  The construction of
\cite{AKOO25} is developed primarily around simple R--NS ramification maps
and is described algebraically in terms of partial super Airy structures.
Our emphasis is instead on keeping all four monodromy assignments explicit
at the operator level and constructing their realisations in super-Miwa variables.

\section{Relation to matrix models}
\label{sec:matrix}
\setcounter{equation}{0}
\subsection{Bosonic case}
Let $u_i,\ i = 1,\ldots N,$ form an orthonormal basis of $\mathbb{C}^N.$ Denoting the $a-$th component of $u_i$ as $u_{i,a}$ we define a (hermitian) matrix $L$ 
with elements
\begin{equation}
L_{ab}= \sum\limits_k l_k u_{k,a} \overline{u_{k,b}}.
\end{equation}
If $f$  is a symmetric function of $l_i,$  then
\begin{equation}
\frac{\partial f}{\partial L_{ab}}
= \sum_{i=1}^n \frac{\partial f}{\partial l_i}\, u_{i,b}\,\overline{u_{i,a}}
\end{equation}
and \[
\frac{\partial^2 f}{\partial L_{ab} \, \partial L_{cd}}
= \sum_{i,j=1}^n 
\frac{\partial^2 f}{\partial l_i \partial l_j}\,
u_{i,b}\,\overline{u_{i,a}}\,
u_{j,d}\,\overline{u_{j,c}}
\;+\;
\sum_{i\neq m}
\frac{\tfrac{\partial f}{\partial l_i}}{l_i - l_m}\,
\Big( u_{m,b}\,\overline{u_{i,a}}\,u_{i,d}\,\overline{u_{m,c}}
+ u_{i,b}\,\overline{u_{m,a}}\,u_{m,d}\,\overline{u_{i,c}} \Big).
\]
In particular, on the space of functions of $\Tr L^m:$
\begin{equation}
\label{eq:second_order_matrix_derivative}
\sum\limits_c 
\frac{\partial^2}{\partial L_{ac} \, \partial L_{cb}}
=
\sum_{i=1}^nu_{i,b}\, \overline{u_{i,a}}
\left(\frac{\partial^2 }{\partial l_i^2} + \sum\limits_{j,j\neq i}\frac{1}{l_i-l_j}\left(\frac{\partial }{\partial l_i}-\frac{\partial }{\partial l_j}\right)\right)
=
\sum_{i=1}^nu_{i,b}\, \overline{u_{i,a}}\ \sT_{\NS}^{\geqslant -1}(l_i),
\end{equation}
with $\sT_{\NS}^{\geqslant -1}(l_i)$ of the form determined in eq.\ (\ref{eq:bosonic_singular_TNS}), and the conditions
\begin{equation} 
\forall\,a,b:\; \sum\limits_c\frac{\partial^2 f}{\partial L_{ac} \, \partial L_{cb}} = 0
\qquad \text{and} \qquad
\forall\, i:\; \sT_{\NS}^{\geqslant -1}(l_i)f = 0
\end{equation}
are equivalent.

Matrix derivatives appear naturally in the Ward identities for the external source matrix integrals of the form
\begin{equation}
Y[L] = \int\! dM\ \mathrm{e}^{-\Tr V(M) + \Tr LM}
\end{equation}
with hermitian $M \in \mathbb{C}^{N\times N}.$ Indeed, under the variation $\delta M = \epsilon M^{m+1}$ we get
\begin{equation}
\label{eq:Matrix_Ward_1}
\int\!DM\, \Tr\!\left(LM^{m+1} - M^{m+1}V'(M) + \sum\limits_{k=0}^m \Tr(M^k)M^{m-k}\right)\ \mathrm{e}^{-\Tr V(M) + \Tr LM} = 0.
\end{equation}
Since 
\begin{equation}
M^i{}_j\mathrm{e}^{\Tr LM} 
=
\frac{\partial}{\partial L^{j}{}_i}\mathrm{e}^{\Tr LM},
\end{equation}
eq.\ (\ref{eq:Matrix_Ward_1}) is equivalent to 
\begin{equation}
\left(\Tr\left(L\partial_L^{m+1}\right) - \Tr\left(\partial_L^{m+1}V'(\partial_L)\right) + \sum\limits_{k=0}^m\Tr(\partial_L)^k\Tr(\partial_L)^{m-k}\right)Y[L] = 0.
\end{equation}
Applying the identity
\begin{equation}
\Tr\left(\partial_L^{m+1}L\right) 
= 
\Tr\left(L(\partial_L)^{m+1}\right)  + \sum\limits_{k=0}^m\Tr(\partial_L)^k\Tr(\partial_L)^{m-k}
\end{equation}
we can  present eq.\ (\ref{eq:Matrix_Ward_1}) in the form
\begin{equation}
    \Tr\left((\partial_L)^{m+1}\left(L-V'(\partial_L)\right)\right)Y[L] = 0.
\end{equation}
 This family of identities is implied by the Gross--Newman equation
\begin{equation}
\label{eq:Gross-Newman}
\left(V'(\partial_L) -L\right)Y[L] = 0.
\end{equation}
Choosing $V(M) = \frac13 M^3$ we conclude that the matrix integral
\begin{eqnarray}
Y[L] = \int\!dM\ \text{e}^{-\frac13\Tr M^3 + \Tr LM}
\label{eq:Kontsevich_matrix_integral}
\end{eqnarray}
with hermitian, $N\times N$ matrices $M, L,$ satisfies for all $i \in \{1,\dots, N\}$ the equation
\begin{equation}
\label{eq:diff_equation_for_YofL}
\left(\frac{\partial^2 }{\partial l_i^2} + \sum\limits_{j,j\neq i}\frac{1}{l_i-l_j}\left(\frac{\partial }{\partial l_i}-\frac{\partial }{\partial l_j}\right) - l_i\right)Y[L] = 0.
\end{equation}
If we rescale the potential by a factor $-\frac{i}{2}$ and define a matrix $\Lambda$ through a relation $L = \frac{i}{2}\Lambda^2$ then we conclude from (\ref{eq:diff_equation_for_YofL}) that
\begin{eqnarray}
Y = \int\!dM\,\exp\,i\Tr\left(\frac16 M^3 + \frac12\Lambda^2 M\right),
\end{eqnarray}
considered as a function of eigenvalues $\lambda_i$ of the matrix $\Lambda,$ satisfies a differential equation
\begin{equation}
\label{eq:diff_equation_for_YofLambda}
\left(\left(\frac{1}{\lambda_i}\frac{\partial}{\partial \lambda_i}\right)^2 
+ 
\sum\limits_{j;\,j\neq i}\frac{2}{\lambda_i^2-\lambda_j^2}
\left(\frac{1}{\lambda_i}\frac{\partial}{\partial \lambda_i} - \frac{1}{\lambda_j}\frac{\partial}{\partial \lambda_j}\right)
-
\lambda_i^2
\right)Y = 0.
\end{equation}
Define now, following \cite{IZ92}
\begin{equation}
Z[\Lambda] = \mathcal{N}_N^{-1} \int\!dM \exp\Bigg\{-\frac12 \Tr\Lambda M^2 + \frac{i}{6}\Tr M^3\Bigg\},
\qquad
\mathcal{N}_N = \int\!dM \exp\Bigg\{-\frac12 \Tr\Lambda M^2\Bigg\}.
\label{eq:norm_Kontsevich_int}
\end{equation}
Since
\begin{equation}
\Tr\Lambda M^2 = \sum\limits_i \lambda_i |M_{ii}|^2 + \sum\limits_{1\leqslant i < j \leqslant N} (\lambda_i + \lambda_j) 
\left((\Re M_{ij})^2 + (\Im M_{ij})^2\right)
\end{equation}
we have
\begin{equation}
\mathcal{N}_N^{-1} =
(2\pi)^{-\frac{N^2}{2}}\prod\limits_{i=1}^N\lambda_i^{\frac12}\hskip -5pt \prod\limits_{1\leqslant i < j \leqslant N}\hskip -5pt (\lambda_i+\lambda_j).
\end{equation}
Shifting $M\to M- i\Lambda$ we get
\begin{equation}
-\frac12 \Tr\Lambda M^2 + \frac{i}{6}\Tr M^3
\to
i\Tr\left(\frac16 M^3 + \frac12\Lambda^2 M\right) + \frac13\Tr\Lambda^3
\end{equation}
so that, up to unimportant constant factor
\begin{equation}
Z[\Lambda] = f^{-1}(\Lambda)\ Y,
\qquad
f(\Lambda) = \prod\limits_{i=1}^N\lambda_i^{-\frac12}\hskip -5pt \prod\limits_{1\leqslant i < j \leqslant N}\hskip -5pt \frac{1}{\lambda_i+\lambda_j}
\exp\Big\{-\tfrac13\sum\limits_j\lambda_j^{3}\Big\}.
\end{equation}
The relation above and the equation (\ref{eq:diff_equation_for_YofLambda}) yield the equation 
satisfied by $Z[\Lambda].$ Denoting
\begin{eqnarray}
\mathcal{L} 
=\left(\frac{1}{\lambda_i}\frac{\partial}{\partial \lambda_i}\right)^2 
+ 
\sum\limits_{j;\,j\neq i}\frac{2}{\lambda_i^2-\lambda_j^2}
\left(\frac{1}{\lambda_i}\frac{\partial}{\partial \lambda_i} - \frac{1}{\lambda_j}\frac{\partial}{\partial \lambda_j}\right)
-
\lambda_i^2
\end{eqnarray}
we have
\begin{equation}
0 = f^{-1}(\Lambda) \mathcal{L}\Big(f(\Lambda) Z[\Lambda]\Big) = \Big(\mathcal{L} + \delta\mathcal{L}\Big)Z[\lambda],
\end{equation}
where
\begin{align}
\label{eq:delta_L_bosonic}
\delta\mathcal{L}
& = 
\frac{2}{\lambda_i^2}
\frac{\partial\log f(\Lambda)}{\partial\lambda_i}
\frac{\partial}{\partial\lambda_i}
 - 
\sum\limits_{j;\,j\neq i}
\frac{2}{\lambda_i^2-\lambda_j^2}
\left(
\frac{1}{\lambda_i}\frac{\partial}{\partial\lambda_i}
-
\frac{1}{\lambda_j}\frac{\partial}{\partial\lambda_j}
\right)
\log f(\Lambda)
\nonumber\\[2pt]
& + 
\frac{1}{\lambda_i^2}
\left[
\frac{\partial^2\log f(\Lambda)}{\partial\lambda_i^2}
+
\left(
\frac{\partial\log f(\Lambda)}{\partial\lambda_i}
\right)^2
\right]
-
\frac{1}{\lambda_i^3}
\frac{\partial\log f(\Lambda)}{\partial\lambda_i}.
\end{align}
It is straightforward to check that
\begin{equation}
\frac{2}{\lambda_i^2}
\frac{\partial\log f(\Lambda)}{\partial\lambda_i}
=
-2\left(
1+
\frac{1}{\lambda_i^2}
\sum_j\frac{1}{\lambda_i+\lambda_j}
\right).
\end{equation}
Computation of the non-derivative terms in (\ref{eq:delta_L_bosonic}) is slightly more involved.
To complete it one uses the identity:
\begin{eqnarray}
\label{eq:the_desired_identity}
\nonumber
\left(\sum\limits_{j}\frac{1}{\lambda_i+\lambda_j}\right)^2
& + & 
\sum\limits_{j,k}\frac{\lambda_i}{\lambda_j}
\left(
\frac{1}{(\lambda_i + \lambda_j)(\lambda_i+\lambda_k)}
+
\frac{1}{(\lambda_i + \lambda_j)(\lambda_j+\lambda_k)}
+
\frac{1}{(\lambda_i+\lambda_k)(\lambda_j+\lambda_k)}
\right)
\\[4pt]
& = &
\left(\sum\limits_j\frac{1}{\lambda_j}\right)^2
\end{eqnarray}
which follows by first noticing that the sum of the third and the fourth (with $j\leftrightarrow k$) terms on the l.h.s.\ of eq.\ (\ref{eq:the_desired_identity}) is equal to
\begin{eqnarray}
\nonumber
&&
\hskip -1cm
\sum\limits_{j,k}\frac{\lambda_i}{\lambda_j}
\left(
\frac{1}{(\lambda_i + \lambda_j)(\lambda_j+\lambda_k)}
+
\frac{1}{(\lambda_i+\lambda_k)(\lambda_j+\lambda_k)}
\right)
=
\sum\limits_{j,k}\left(\frac{1}{\lambda_j} + \frac{1}{\lambda_k}\right)\frac{\lambda_i}{(\lambda_i + \lambda_j)(\lambda_j+\lambda_k)}
\\[4pt]
& = &
\sum\limits_{j,k}\frac{\lambda_i}{(\lambda_i+\lambda_j)\lambda_j\lambda_k}
=
\sum\limits_j\frac{1}{\lambda_j}\sum\limits_k\frac{\lambda_i}{\lambda_k(\lambda_i+\lambda_k)},
\end{eqnarray}
while the sum of the first and the second term reads
\begin{equation}
\sum\limits_k\frac{1}{\lambda_i+\lambda_k}\sum\limits_j\frac{\lambda_i}{\lambda_i+\lambda_j}\left(\frac{1}{\lambda_i}+\frac{1}{\lambda_j}\right)
=
\sum\limits_j\frac{1}{\lambda_j} \sum\limits_k\frac{1}{\lambda_i+\lambda_k}.
\end{equation}
Adding up this terms we get
\begin{equation}
\sum\limits_j\frac{1}{\lambda_j}\sum\limits_k\frac{1}{\lambda_i+\lambda_k}\left(\frac{\lambda_i}{\lambda_k} + 1\right)
=
\left(\sum\limits_j\frac{1}{\lambda_j}\right)^2
\end{equation}
as desired. An operator annihilating $Z[\Lambda]$ thus has the form
\begin{equation}
\left(\frac{1}{\lambda_i}\frac{\partial}{\partial \lambda_i}\right)^2 
+ 
\sum\limits_{j;\,j\neq i}\frac{2}{\lambda_i^2-\lambda_j^2}
\left(\frac{1}{\lambda_i}\frac{\partial}{\partial \lambda_i} - \frac{1}{\lambda_j}\frac{\partial}{\partial \lambda_j}\right)
-2\left(1+\frac{1}{\lambda_i^2}\sum\limits_{j}\frac{1}{\lambda_i+\lambda_j}\right)\frac{\partial}{\partial\lambda_i}
+
\frac{1}{\lambda_i^2}\left(\sum\limits_j\frac{1}{\lambda_j}\right)^2 + \frac14\frac{1}{\lambda_i^4}.
\end{equation}
Changing variables to $l_i = \lambda_i^2$ and comparing with  eq.\ (\ref{eq:Airy_structure_R_bosonic}) we conclude, that $Z[\Lambda]$ is the partition function
for the bosonic, Ramond Airy structure.

The relevance of this result stems from the fact, that normalized Kontsevich integral
\eqref{eq:norm_Kontsevich_int} admits a direct geometric
interpretation. Evaluating it perturbatively with the quadratic  in $M$ term defining the ``free action'' and the cubic term
the ``interaction'', we get the propagator 
\begin{equation}
\left\langle M_{ij}M_{kl}\right\rangle
=
\frac{2}{\lambda_i+\lambda_j}
\delta_{il}\delta_{jk}.
\label{eq:Kontsevich-matrix-propagator}
\end{equation}
The two Kronecker deltas produce the double-line structure of the
propagator, while the cubic interaction produces trivalent vertices
\cite{BIZ80,IZ79}. The double-line notation fixes a cyclic ordering at every vertex of the graph, so that each diagram can be thickened to a compact oriented surface. The normalization $\mathcal{N}_{N}$ removers a Gaussian vacuum factors (diagrams), and taking logarithm choose only connected diagrams.  Therefore, the perturbative expansion of $\log Z[\Lambda]$ is organised by connected trivalent ribbon graphs \cite{Kontsevich92}.  We
suppress throughout the standard rescaling of $M$ which absorbs the factors
of $i$ accompanying the cubic vertices. The edge factor in \eqref{eq:Kontsevich-matrix-propagator} has the Laplace
representation \cite{GL24}
\begin{equation}
\frac{2}{\lambda_i+\lambda_j}
=
2\int_0^\infty \!d\ell_e\,
e^{-\ell_e(\lambda_i+\lambda_j)},
\label{eq:Kontsevich-edge-Laplace}
\end{equation}
so that every edge $e$ acquires a positive length $\ell_e$ and the Feynman
diagrams become metric ribbon graphs.  For a face $f$ carrying the matrix index $i$, the exponential factors associated with its boundary combines as
\begin{equation}
\prod_{e\subset\partial f}e^{-\lambda_i\ell_e}
=
e^{-\lambda_i p_f},
\qquad
p_f
=
\sum_{e\subset\partial f}\ell_e,
\label{eq:Kontsevich-face-perimeter}
\end{equation}
where edge incidences are counted with their multiplicities.  Thus $p_f$ is
the total boundary length, or perimeter, of the face, and $\lambda_i$ is
its Laplace-conjugate variable.  The graph dual to a trivalent ribbon graph
gives a possibly degenerate triangulation of the thickened surface.  Its
vertices correspond to the faces of the ribbon graph and hence to the
marked points.

The matrix size $N$ fixes the range of face label, unlike in an usual Hermitian matrix models, it does not automatically grade the expansion by genus $g$. Instead, $g$ is determined by standard Euler characteristic $V-E+F=2-2g$ with $F=n$, where $V$, $E$, and
$F$ denote the numbers of vertices, edges, and faces of the ribbon graph respectively, where 
faces also  are in one-to-one correspondence with the $n$  marked points on Riemann surface.
The metric ribbon graphs appearing here are related to decorated Riemann
surfaces through Strebel quadratic differentials \cite{Strebel1984}.
Kontsevich \cite{Kontsevich92} showed that these graphs label the cells
of the combinatorial moduli space
\begin{equation}
\mathcal M_{g,n}^{\mathrm{comb}}
\simeq
\mathcal M_{g,n}\times\mathbb R_+^n,
\label{eq:combinatorial-moduli}
\end{equation}
where the factor $\mathbb R_+^n$ records the $n$ face perimeters
$p_1,\ldots,p_n$ introduced in
\eqref{eq:Kontsevich-face-perimeter}.  The matrix-model
weights reproduce the weights of the corresponding cells, including their
automorphism factors, and the resulting graph sum computes the intersection
numbers of stable curves entering the Kontsevich--Witten partition function.  Contracting
edges of vanishing length describes the passage to boundary cells.  The
relation between this graph compactification and the appropriate quotient
of the Deligne--Mumford compactification $\overline{\mathcal M}_{g,n}$ was
made precise in \cite{Zvonkine04}.

This construction also closes the circle with the Miwa variables of
Section~\ref{sec: Section 3}.  After the sum over face labels, the
dependence on the external matrix is organised by
\begin{equation}
g_k
=
\frac{1}{k}\sum_{j=1}^N l_j^{-k}
=
\frac{1}{k}\Tr\Lambda^{-2k},
\qquad
l_j=\lambda_j^2,
\qquad
k\in\mathbb Z_{\geqslant0}+\frac12.
\label{eq:Ramond-Miwa-Kontsevich-times}
\end{equation}
For $k=m+\frac12$, this gives
\begin{equation}
g_{m+\frac12}
=
\frac{2}{2m+1}
\Tr\Lambda^{-(2m+1)}.
\end{equation}
Thus, after passing to the coordinate $l=\lambda^2$, the odd Miwa times of
the Kontsevich model are naturally identified, up to the conventional
normalization and sign, with the half-integer times of our bosonic Ramond
current.  The half-integer moding is that of $\jmath_{\R}$, that is, of a
bosonic current antiperiodic around the branch point of $x=z^2$; the
identification therefore depends on the choice of the square-root coordinate
and does not assign an intrinsic Ramond sector to the original bosonic
matrix model.  The matrix integral is thus not merely a source of the
differential constraints above,  its perturbative expansion provides a
combinatorial description of the geometry of the moduli space.

In the present work this bosonic, well known correspondence serves as a
motivation and a guideline.  Whether an analogous combinatorial interpretation exists on
the super-Miwa side is as an open problem.

\subsection{The super-prefactor}

We will analyze in this sub-section the RR super-Airy structure along the lines which allowed to relate the bosonic, Ramond
Airy structure to the hermitian matrix model and briefly comment on the R-NS and NS-R cases.

Recall that (c.f.\ eq.\ (\ref{eq:final_expression_for_TRR}))
\begin{equation}
\mT^{\geqslant-\frac12}_{\R\R}(l_i,\theta_i) = \mT^{\geqslant-\frac12}_{\NS\NS}(l_i,\theta_i) + 
\delta\mT^{\geqslant-\frac12}(l_i,\theta_i).
\end{equation}
Substituting $l_j = \lambda_j^2$ we get 
\begin{eqnarray}
\nonumber
4\mT^{\geqslant-\frac12}_{\NS\NS}(\lambda^2_i,\theta_i)
& = &
\theta_i\left(\frac{1}{\lambda_i}\frac{\partial}{\partial\lambda_i}\right)^2
+
 \frac{2}{\lambda_i}\frac{\partial}{\partial\lambda_i}\frac{\partial}{\partial \theta_i}
+ 
\sum\limits_{j;\,j\neq i}\frac{2\theta_i-\theta_j}{\lambda_i^2-\lambda_j^2}\left(\frac{2}{\lambda_i}\frac{\partial}{\partial\lambda_i} - \frac{2}{\lambda_j}\frac{\partial}{\partial\lambda_j}\right)
\\[4pt]
& + &
\sum\limits_{j;\,j\neq i}\frac{4}{\lambda^2_i-\lambda^2_j-\theta_i\theta_j}\left(\frac{\partial}{\partial\theta_i} - \frac{\partial}{\partial\theta_j}\right)
\end{eqnarray}
and
\begin{eqnarray}
&&
\hskip -1cm
4\delta \mT^{\geqslant-\frac12}(\lambda^2_i,\theta_i) 
=
-
\frac{4}{\lambda_i}\sum\limits_j \frac{1}{\lambda_i + \lambda_j}\left(\frac{\theta_i}{\lambda_i} \frac{\partial}{\partial \lambda_i} + \frac{\partial}{\partial \theta_i}\right)
-
\frac{1}{\lambda_i^2}\left(\sum\limits_{j;\,j\neq i}\frac{\theta_j}{\lambda_j}\frac{\lambda_i-\lambda_j}{\lambda_i+\lambda_j}\right)\frac{\partial}{\partial\lambda_i}
\\[4pt]
\nonumber
& + & 
\frac{1}{\lambda_i}\sum\limits_{j;j\neq i}
\frac{\theta_j}{\lambda_i^2-\lambda_j^2}\Big(\frac{2}{\lambda_i+\lambda_j} - \frac{1}{\lambda_j}\Big)\theta_i\frac{\partial}{\partial\theta_i}
+
\frac{\theta_i}{\lambda_i^4}+ \frac{\theta_i}{\lambda_i^2}\left(g_{\frac12}^2 + 2q_1q_0\right) + \frac{1}{\lambda_i^2}q_0 g_\frac12 
\\[4pt]
\nonumber
& - & 
\sum_{j\neq i}
\frac{2\lambda_i+\lambda_j}
{\lambda_i^3(\lambda_i+\lambda_j)^2}
\theta_i\theta_j
\frac{\partial}{\partial\theta_i} .
\end{eqnarray}
Notice first, that conjugating $4\mT^{\geqslant-\frac12}_{\NS\NS}(\lambda^2_i,\theta_i) $ by the factor of the form\footnote{
We note that the fermionic part of $F_0$ is precisely the difference of the NS and R fermionic two-point functions (Szeg\H{o} kernels), evaluated at the Miwa points,   $F_0 \sim \exp\left[ \sum_{j<k}\theta_j\theta_k \Big( \langle\psi_{\NS}(l_j)\psi_{\NS}(l_k)\rangle - \langle\psi_{\R}(l_j)\psi_{\R}(l_k)\rangle \Big) \right], $ since $\langle\psi_{\NS}(x)\psi_{\NS}(y)\rangle=(x-y)^{-1}$ and $\langle\psi_{\R}(x)\psi_{\R}(y)\rangle =\tfrac{x+y}{2\sqrt{x}\sqrt{y}}\,(x-y)^{-1}$ give, at $x=\lambda_j^2$, $y=\lambda_k^2$,  $ \langle\psi_{\NS}\psi_{\NS}\rangle-\langle\psi_{\R}\psi_{\R}\rangle = -\frac{\lambda_j-\lambda_k}{2\lambda_j\lambda_k(\lambda_j+\lambda_k)}\ $.}
\begin{equation}
\label{eq:F_0_factor}
F_0 = \prod\limits_j\lambda_j^{-1}\hskip -5pt \prod\limits_{j,k;\,j < k}(\lambda_j+\lambda_k)^{-2}
\exp\Big\{-\frac{1}{2}\sum\limits_{j,k;\,j < k}\frac{\theta_j\theta_k}{\lambda_j\lambda_k}\frac{\lambda_j-\lambda_k}{\lambda_j+\lambda_k}\Big\},
\end{equation}
and retaining for now only these of the resulting terms which contain derivatives,
\begin{eqnarray}
&&
\hskip -2.5cm
F_0^{-1}\left(\theta_i\left(\frac{1}{\lambda_i}\frac{\partial}{\partial\lambda_i}\right)^2 + 
\frac{2}{\lambda_i}\frac{\partial}{\partial\lambda_i}\frac{\partial}{\partial\theta_i}\right)F_0
- \theta_i\left(\frac{1}{\lambda_i}\frac{\partial}{\partial\lambda_i}\right)^2 
- \frac{2}{\lambda_i}\frac{\partial}{\partial\lambda_i}\frac{\partial}{\partial\theta_i}
\\[4pt]
\nonumber
& = &
\left(\frac{2\theta_i}{\lambda_i^2}\frac{\partial\log F_0}{\partial\lambda_i} -\frac{\theta_i}{\lambda_i^3}
+
\frac{2}{\lambda_i}\frac{\partial\log F_0}{\partial\theta_i}\right)\frac{\partial}{\partial\lambda_i}
+
\frac{2}{\lambda_i}\frac{\partial\log F_0}{\partial\lambda_i}\frac{\partial}{\partial\theta_i}
+
\ldots
\end{eqnarray}
we reproduce all terms in $4\delta \mT^{\geqslant-\frac12}(\lambda^2_i,\theta_i)$  containing derivatives.

Computation of the remaining (non-derivative) terms in $F_0^{-1}4\delta \mT^{\geqslant-\frac12}(\lambda^2_i,\theta_i)F_0$
is tedious, but straightforward. 

Concentrating first on terms linear in $\theta_s,$ and using identity
\begin{eqnarray}
& - & 
\sum\limits_{j,k}\frac{2\theta_j}{\lambda_i}\left(\frac{1}{\lambda_j(\lambda_i+\lambda_j)(\lambda_i+\lambda_k)} \
+ \frac{1}{\lambda_j(\lambda_i+\lambda_j)(\lambda_j+\lambda_k)} 
+\frac{1}{\lambda_j(\lambda_i+\lambda_k)(\lambda_j+\lambda_k)}\right)
\\[2pt]
\nonumber
& - &
\sum\limits_{j,k}\frac{2\theta_j}{\lambda_i}\left(\frac{1}{\lambda_k(\lambda_i+\lambda_j)(\lambda_i+\lambda_k)} \ 
+\frac{1}{\lambda_k(\lambda_i+\lambda_j)(\lambda_j+\lambda_k)} 
+\frac{1}{\lambda_k(\lambda_i+\lambda_k)(\lambda_j+\lambda_k)}\right)
\\[2pt]
\nonumber
& - &
\sum\limits_{j,k} \frac{4\theta_j}{\lambda_i}\frac{1}{\lambda_i(\lambda_i+\lambda_j)(\lambda_i+\lambda_k)}
=
-\frac{4}{\lambda_i^2}\sum\limits_j\frac{\theta_j}{\lambda_j}\sum\limits_k\frac{1}{\lambda_k}
\end{eqnarray}
we arrive at the expression
\begin{equation}
\frac{4\theta_i}{\lambda_i^2} \left(\sum\limits_j\frac{1}{\lambda_j}\right)^2
+ \frac{\theta_i}{\lambda_i^4}
-\frac{2}{\lambda_i^2}\sum\limits_j\frac{\theta_j}{\lambda_j}\sum\limits_k\frac{1}{\lambda_k}.
\end{equation}
Since
\begin{equation}
\sum\limits_j\frac{\theta_j}{\lambda_j} = -q_0, \qquad \sum\limits_j\frac{1}{\lambda_j} = \frac12g_{\frac12},
\end{equation}
we can rewrite this result as
\begin{equation}
\frac{\theta_i}{\lambda_i^4}
+
\frac{1}{\lambda_i^2}\left(\theta_i g_{\frac12}^2 + q_0g_\frac12\right).
\end{equation}

Terms containing three $\theta-$s 
come from
\begin{equation}
\frac{2}{\lambda_i}\frac{\partial \log f_\theta}{\partial\lambda_i} \frac{\partial \log f_\theta}{\partial\theta_i},
\qquad
\sum\limits_{j;\,j\neq i}\frac{2\theta_i-\theta_j}{\lambda_i^2-\lambda_j^2}
\left(
\frac{2}{\lambda_i}\frac{\partial  \log f_\theta}{\partial\lambda_i} - \frac{2}{\lambda_j}\frac{\partial  \log f_\theta}{\partial\lambda_j}\right)
\label{eq:sing}
\end{equation}
and
\begin{equation}
\sum\limits_{j;\,j\neq i}\frac{4\theta_i\theta_j}{(\lambda_i^2-\lambda_j^2)^2}
\left(
\frac{\partial  \log f_\theta}{\partial\theta_i} -\frac{\partial  \log f_\theta}{\partial\theta_j}\right),
\end{equation}
where 
\begin{eqnarray}
 f_\theta = \exp\Bigg\{-\frac{1}{2}\sum\limits_{j,k;\,j < k}\frac{\theta_j\theta_k}{\lambda_j\lambda_k}\frac{\lambda_j-\lambda_k}{\lambda_j+\lambda_k}\Bigg\}
\end{eqnarray}
is the exponential term present in (\ref{eq:F_0_factor}). 
To compute their sum (separately, these terms seem not to have a simple form) it is useful to first present all the double sums, with summation indices, say, $j$ and $k,$ in a form such that the range of $j$ and $k$ is not restricted, by explicitly adding diagonal terms with $j=i$ and/or $k =i$ (this is possible since the powers of $\lambda_i-\lambda_{j,k}$ in denominators cancel against powers of $\lambda_i-\lambda_{j,k}$ arising from explicitly computing derivatives of $ \log f_\theta$). Then, for a sum of the form
\begin{equation}
\sum\limits_{j,k}\theta_j\theta_k h_{jk},
\end{equation}
only anti-symmetric part of the $h_{jk}$ survives. A an intermediate step of the calculation we obtain an expression
\begin{equation}
\sum\limits_{j,k}\frac{\theta_i\theta_j\theta_k}{\lambda_i\lambda_k}\left(
\frac{1}{\lambda_i\lambda_j(\lambda_i+\lambda_j)^2}
+
\frac{1}{\lambda_i^2\lambda_j(\lambda_i+\lambda_k)} 
+
\frac{2}{\lambda_i\lambda_j^3} + \frac{1}{\lambda_i^2(\lambda_i+\lambda_j)^2} \right).
\end{equation}
Since 
\begin{equation}
\frac{1}{\lambda_i\lambda_j(\lambda_i+\lambda_j)^2} + \frac{1}{\lambda_i^2(\lambda_i+\lambda_j)^2}
=
\frac{1}{\lambda_i^2\lambda_j(\lambda_i+\lambda_j)}
\end{equation}
and
\begin{equation}
\sum\limits_{j,k}\frac{\theta_i\theta_j\theta_k}{\lambda_i^3\lambda_j\lambda_k}\left(\frac{1}{\lambda_i+\lambda_j} + \frac{1}{\lambda_i+\lambda_k}\right) = 0,
\end{equation}
we are left with
\begin{equation}
\frac{2\theta_i}{\lambda_i^2}\sum\limits_j\frac{\theta_j}{\lambda_j^3}\sum\limits_k\frac{\theta_k}{\lambda_k}
=
\frac{2\theta_i}{\lambda_i^2} q_1q_0.
\end{equation}
We thus arrive at the strikingly simple formula
\begin{equation}
\mT^{\geqslant-\frac12}_{\R\R}(l_i,\theta_i)
=
F_0^{-1}\mT^{\geqslant-\frac12}_{\NS\NS}(l_i,\theta_i)F_0.
\end{equation}

As a direct corollary, if $Z_{\NS\NS}$ satisfies the NS--NS constraints
on the Miwa locus, then
\begin{equation}
Z_{\R\R}
=
F_0^{-1}Z_{\NS\NS}
\label{eq:NSNS-RR-partition-functions}
\end{equation}
satisfies the corresponding R--R constraints, since
\begin{equation}
\mT^{\geqslant-\frac12}_{\R\R}Z_{\R\R}
=
F_0^{-1}
\mT^{\geqslant-\frac12}_{\NS\NS}Z_{\NS\NS}
=
0.
\end{equation}
Thus the two differential constraint systems are equivalent on the Miwa
locus; this also accounts for the half-integer moding of the projected odd
R--R generators in this realisation.  Before imposing the parity and
normalisation conditions described above, the R--R solution is understood
up to the residual dependence on the additional odd coordinate $q_0$.
Equation~\eqref{eq:NSNS-RR-partition-functions} does not assert an
equivalence of the partition functions on the full abstract space of
times.  In particular, taking $Z_{\NS\NS}=1$ reproduces
$\mT^{\geqslant-\frac12}_{\R\R}F_0^{-1}=0$.

Together with the cubic Airy weight we now introduce the full prefactor
\begin{equation}
    F=F_{0}\exp \left(-\frac{\mu}{3}\sum\limits_j \lambda_j^3 \right).
    \label{eq:full-prefactor}
\end{equation}
Direct conjugation by \eqref{eq:full-prefactor}, before adding
the Ramond zero-mode contribution, gives
\begin{equation}
F^{-1}\left(4\mT_{\NS\NS}^{\geqslant-\frac12}(\lambda_i^2,\theta_i)
-
\mu^2\theta_i\lambda_i^2
+
\mu q_0\right)F
=
4\mT_{\R\R}^{\geqslant-\frac12}(\lambda_i^2,\theta_i)
-
2\mu \left(\theta_i\frac{\partial}{\partial\lambda_i}
+
\lambda_i\frac{\partial}{\partial\theta_i} \right).
\label{eq:NSNS-RR-intertwiner-without-zero-mode}
\end{equation}

To match the normalisation of the cubic factor in $F$, throughout this
subsection we use the R--R Airy operator
\eqref{eq:Airy_structure_RR} with $\mu$ replaced by $\mu/2$.  Its linear
contribution in the original Miwa variables is
\begin{equation}
\left.
\delta\mT_{\R\R}^{\geqslant-\frac12}(l_i,\theta_i)
\right|_{k=\frac32,\mathrm{lin}}
=
-\frac{\mu}{2}\sqrt{l_i}
\left(
D_i+\theta_iD_i^2
\right)
-
\frac{\mu}{2}
\frac{\partial}{\partial q_0}.
\label{eq:RR-Heisenberg-Airy-shift}
\end{equation}
Equivalently, after $l_i=\lambda_i^2$ and multiplication by $4$, define
\begin{equation}
{4\mT}^{\geqslant  - \frac12}_{\R\R,\Ai}(\lambda_i^2,\theta_i)
=
4\mT_{\R\R}^{\geq-\frac12}(\lambda_i^2,\theta_i)
-
2\mu
\left(
\theta_i\frac{\partial}{\partial\lambda_i}
+
\lambda_i\frac{\partial}{\partial\theta_i}+\frac{\partial}{\partial q_0}
\right)
\label{eq:RR-Airy-operator-matrix-section}
\end{equation}
The derivative $\partial_{q_0}$ removes the $m=0$ term contained in the
local Miwa identity: the latter resums the Ramond fermionic current starting
with $m=0$, whereas the projected cubic shift starts with $\psi_1$.
For arbitrary finite $N$, assuming that $\lambda_j\neq0$ and
$\lambda_j^2\neq\lambda_r^2$ for $j\neq r$, and fixing the branches
$\lambda_j=\sqrt{l_j}$, the multiplicative inverse of the fermionic Miwa
transformation, with $q_m$, $1\leq m\leq N-1$, held fixed, gives
\begin{equation}
q_0
=
-\sum_j\frac{\theta_j}{\lambda_j},
\qquad
\frac{\partial}{\partial q_0}
=
-\sum_j
c_j\lambda_j
\frac{\partial}{\partial\theta_j},
\qquad
c_j
=
\prod_{r\neq j}
\left(
1-\frac{\lambda_r^2}{\lambda_j^2}
\right)^{-1}.
\label{eq:finite-Ramond-zero-mode}
\end{equation}
Consequently, the Ramond zero mode is represented by
\begin{equation}
\psi_0
=
\frac{q_0}{2}
+
\frac{\partial}{\partial q_0}
=
-\frac12\sum_j\frac{\theta_j}{\lambda_j}
-
\sum_j
c_j\lambda_j
\frac{\partial}{\partial\theta_j}.
\label{eq:Ramond-zero-mode-Miwa}
\end{equation}

At the level of the abstract R--R Airy structure, $q_0$ is an
independent additional odd coordinate.  For arbitrary finite $N$,
equations \eqref{eq:finite-Ramond-zero-mode} and
\eqref{eq:Ramond-zero-mode-Miwa} give its Miwa realisation.  On the Miwa
locus, $q_0$ is pulled back to a linear combination of the Grassmann
Miwa variables $\theta_j$, while $\partial_{q_0}$ is represented by the
corresponding vector field in the $\theta_j$ variable. At finite $N$, this vector field keeps fixed the fermionic times $q_m$ only
for $1\leq m\leq N-1$; in general it does not annihilate the higher times
$q_m$ with $m\geq N$.
Thus, these formulas are
identities on the Miwa locus rather than identifications in the full
abstract Weyl algebra.  Their $N\to\infty$ extension is understood as
the corresponding formal limit.
In the Miwa variables, the Grassmann part of $F$ depends on the variables $\theta_j$.  Consequently, multiplication by $F$ does not
commute with the vector field representing $\partial_{q_0}$ in
\eqref{eq:finite-Ramond-zero-mode}.  Acting with this vector field gives
\begin{equation}
\frac{\partial}{\partial q_0}\log F
=
\frac12
\sum_{j,k}
c_j\theta_k
\frac{\lambda_j-\lambda_k}
{\lambda_k(\lambda_j+\lambda_k)}
=  -\frac{q_0}{2} - \sum_{j,k}
\frac{c_j\theta_k}{(\lambda_j+\lambda_k)} =
-\frac{q_0}{2}
-
S(\lambda,\theta),
\label{eq:q0-log-F}
\end{equation}
where we 
denote 
$
S(\lambda,\theta)
=
\sum_{j,k}
c_j\theta_k/(\lambda_j+\lambda_k)
$.
In the second equality we used $\sum_jc_j=1$ together with
$
(\lambda_j-\lambda_k)/
[\lambda_k(\lambda_j+\lambda_k)]
=
\lambda_k^{-1}-2(\lambda_j+\lambda_k)^{-1}
$.
Since $\psi_0=q_0/2+\partial_{q_0}$, equation
\eqref{eq:q0-log-F} implies
$
F^{-1}\psi_0F
=
\partial_{q_0}-S(\lambda,\theta)
$.
Thus the bare Ramond zero mode is not conjugated to
$\partial_{q_0}$ alone.

Combining this result with
\eqref{eq:NSNS-RR-intertwiner-without-zero-mode}, we obtain
\begin{equation}
\begin{aligned}
F^{-1}
\bigg[
4\mT_{\NS\NS}^{\geq-\frac12}(\lambda_i^2,\theta_i)
-\mu^2\theta_i\lambda_i^2
+\mu q_0
-2\mu\psi_0
\bigg]
F
=
{4\mT}^{\geqslant  - \frac12}_{\R\R,\Ai}(\lambda_i^2,\theta_i)
+
2\mu S(\lambda,\theta).
\end{aligned}
\label{eq:NSNS-RR-similarity-explicit}
\end{equation}
The additional, odd operator is therefore a multiplicative
contribution produced by the Grassmann part of the prefactor.
Equivalently, we may introduce the covariant Ramond zero-mode operator
$
\nabla_{q_0}
=
\psi_0+S(\lambda,\theta)
=
\partial_{q_0}-\partial_{q_0}\log F
$.
By construction,
$
F^{-1}\nabla_{q_0}F=\partial_{q_0}
$,
and the intertwining relation takes the exact form
\begin{equation}
\begin{aligned}
F^{-1}
\bigg[
4\mT_{\NS\NS}^{\geq-\frac12}(\lambda_i^2,\theta_i)
-\mu^2\theta_i\lambda_i^2
+\mu q_0
-2\mu\nabla_{q_0}
\bigg]
F
=
{4\mT}^{\geqslant  - \frac12}_{\R\R,\Ai}(\lambda_i^2,\theta_i).
\end{aligned}
\label{eq:NSNS-RR-similarity}
\end{equation}
Equations
\eqref{eq:NSNS-RR-similarity-explicit} and
\eqref{eq:NSNS-RR-similarity} are equivalent.  The first keeps the
additional odd multiplication operator explicit, whereas the second
absorbs it into the covariant Ramond zero mode.  In this form the
similarity transformation separates the change of sector, generated by
$F_0$, from the cubic Airy deformation contained in $F$.

The derivatives
$D_i=\partial_{\theta_i}+\theta_i\partial_{l_i}$
appearing throughout Section~\ref{sec: Section 3} are superderivatives on
the space of spectral variables.  Equation
\eqref{eq:second_order_matrix_derivative} provides a genuine
matrix-derivative interpretation of their bosonic part. The relations derived above should therefore be
understood as intertwining identities between
the NS--NS and R--R differential constraints. The
relations derived above are therefore intertwining identities between the
NS--NS and R--R differential constraints on the Miwa locus.  They are not
an isomorphism of the abstract super-Virasoro representations: the
Ramond-sector objects $q_0$, $\psi_0$ and $\nabla_{q_0}$ have no NS--NS
counterpart, and the rational prefactors $F_0$ and $F$ need not define
formal power series in the full set of abstract times.

\section{Discussion and outlook}
\label{sec:outlook}
In this work, we have established a unified supercurrent framework that effectively linearises the $\mathcal{N}=1$ super-Virasoro constraints across the full range of Neveu–Schwarz and Ramond monodromy assignments. By constructing sector-dependent super-Miwa transformations, we have provided an explicit differential realization of these constraints, culminating in an exact similarity transformation that intertwines the NS–NS and R–R sectors on the Miwa locus through a specific super-Kontsevich prefactor. This identity not only clarifies the interplay between different fermionic boundary conditions in the spectral variable representation but also supplies the foundational Ward identities for any prospective supersymmetric extension of the Kontsevich model. Consequently, our results delineate the precise algebraic and analytic requirements for a matrix-model measure capable of recovering these superconformal structures, thereby bridging the gap between abstract super-Airy constraints and their potential combinatorial realization. This foundation opens the path toward addressing several open problems, particularly regarding the geometric interpretation of the resulting partition functions and the inclusion of more general polarization backgrounds.

The single super-current $\mJ(x,\vartheta)$ on super-surfaces has served here as more than a notational device.  Since the projected tensor is quadratic in one object, 
the four monodromy assignments are handled by the same residue construction, and a single dilaton shift of the bosonic current produces
the linear terms of the Virasoro and of the super-Virasoro generators
simultaneously.  The price is that the sectors are distinguished only by
the moding of $\mJ$, so that features which are conceptually distinct are 
the Ramond fermionic zero mode, the shift of $L_0$ by $-\tfrac1{16}$, the
presence or absence of $G_0$,  all enter through one and the same
formula and have to be disentangled afterwards, as in
Section~\ref{sec:airy}.  The same economy is visible in the super-Miwa
representation: the defining relation $\mJ_>(l_i,\theta_i)=-D_i$ fixes the
bosonic and fermionic times together, and the two-body kernel
$(l_i-l_j-\theta_i\theta_j)^{-1}$ appears as a single object rather than
as a Bergman and a Szeg\H{o} contribution added by hand into the theory.

The results of this paper are systems of constraints rather than
recursions.  The comparison with the $\mathcal N=1$ super-topological
recursion of \cite{BO21} and with the superconformal topological
recursion of \cite{AKOO25} was made in Section~\ref{sec:airy} at the
level of the local quadratic density, which is where the super-current
tensor and the super quadratic Casimir can be compared directly.
Establishing the correspondence at the level of the recursion itself
would require constructing the multilinear differentials $\omega_{g,n}$
generated by the constraints derived here and matching them against the
super loop equations.  In the local NS--NS setting with trivial
polarisation this is largely a matter of translating mode and covering
conventions.  In the sectors containing a Ramond fermion the situation is
different, since the zero mode has no direct counterpart on the
recursion side, and the super-Miwa representation of $\psi_0$ obtained in
Section~\ref{sec:matrix} may provide a useful starting point.

The four sectors are not on an equal footing as far as the fermionic zero
mode is concerned, and it is worth recording why.  In the NS--NS and
R--NS sectors the fermion is of Neveu--Schwarz type and no fermionic zero
mode occurs, so that the dilaton shift produces an ordinary Super Quantum
Airy Structure.  In the R--R sector the modes of the super-stress tensor
are half-integer, the odd generators carry no $G_0$, and $q_0$ enters as
an additional odd coordinate.  The NS--R sector is the only one in which
the Ramond fermionic zero mode $\psi_0=\tfrac{q_0}{2}+\partial_{q_0}$ and
the odd generator $G_0$ are present simultaneously; there the choice of
the dilaton shift and the choice of the subalgebra imposed on the
partition function cannot be made independently, and the two realisations
described in Section~\ref{sec:airy} differ precisely in how this is
resolved.  Both lead to the same conclusion, one additional odd
coordinate $q_0$, but they are not equivalent as constructions: one
retains the full family $L_n,G_n$ at the cost of an auxiliary constraint
for the bosonic zero mode, the other removes $G_0$ and leaves the
constant in $L_0$ undetermined.  A geometric criterion selecting between
them, presumably on the side of the super-spectral curve, would settle
the question.

The four sector-dependent super-Miwa realisations derived in this work,
together with the Ramond zero-mode contribution and the prefactor $F_0$
intertwining the NS--NS and R--R sectors, provide a concrete system of
Ward identities that a supersymmetric analogue of the Kontsevich model
would have to reproduce.  At present, however, we do not interpret these operators as radial
operators of an established supermatrix integral.  Several related
constructions have been considered in the literature.  Standard
$c$-type Hermitian supermatrix models depend on the matrix dimensions
only through the superdimension $N-M$ and therefore reduce formally to
ordinary bosonic matrix models \cite{AM91,Yost92}.  This reduction does
not produce the fermionic radial operators required by the super-Miwa
representation.
Super-eigenvalue models instead realise super-Virasoro constraints
directly in bosonic and Grassmann eigenvalue variables
\cite{Itoyama91,Mikhailov93,ABBEM93,BB93,AIMZ92,
Ciosmak:2016wpx,Ciosmak:2017ofd}, but they are not obtained as radial
reductions of an external-source supermatrix integral.  External-source
supermatrix models have been investigated in
\cite{Kimura14,Kimura23}, while a different construction based on
coupled bosonic and fermionic matrix ensembles was proposed in
\cite{Makeenko96}; see also \cite{Takama92} for a related matrix-model
realisation lifted from  super-eigenvalue Virasoro constraints.  None of these constructions
produces an external-source integral whose radial Ward identities
reproduce the four sector-dependent super-Miwa operators derived here,
including their Ramond zero-mode realisation.
 
Any future matrix model would have to reproduce several specific features
of the present construction.  Its measure should generate the bosonic
factor $\prod_i\lambda_i^{-1}\prod_{i<j}(\lambda_i+\lambda_j)^{-2}$
which, up to an eigenvalue-independent constant, is the square of the
Gaussian normalisation $\mathcal N_N$ of \eqref{eq:norm_Kontsevich_int};
its fermionic part should reproduce the Grassmann kernel of $F_0$, given
by the difference between the Neveu--Schwarz and Ramond fermionic
propagators.  It should also incorporate the Ramond zero mode and admit a
radial or HCIZ-type reduction leading to the super-Miwa variables
$(l_i,\theta_i)$.  Whether a model with these properties can be
constructed, and whether its perturbative expansion admits a ribbon-graph
interpretation related to the geometry of supermoduli space, remains
open.
 
A further natural extension concerns the polarisation.  The explicit
operators obtained here correspond to $\varphi_{mn}=\chi_{rs}=0$, that is,
to the polarisation inherited from $\mathbb{CP}^{1}$, while the general
deformation was only formulated at the level of the state at infinity.
Reinstating $\varphi_{mn}$ and $\chi_{rs}$ dresses the
two-body kernel by the regular parts of the Bergman and Szeg\H{o}
kernels, and the resulting operators would be the natural super-Miwa
counterpart of a global spectral curve.  When negative modes are included
the dilaton shift and the change of polarisation no longer commute, and
the additional terms generated in this way, as in \cite{AKOO25}, deserve
a separate analysis in the present framework.

{\bf{Acknowledgments} } \\
This work was supported by the Polish National Science Centre (NCN) under grant\\  No.\ 2023/49/B/ST2/03481. 
LH would like to thank the organizers of the 2026 Simons Physics Summer Workshop and the Simons Center for Geometry and Physics 
for their hospitality and for providing a stimulating environment in which this work was completed, and Ond\v{r}ej Hul\'ik for numerous discussions.

\sloppy

\bibliographystyle{JHEP} 

\bibliography{bibliography}

\newpage

\end{document}